\documentclass[journal,draftcls,onecolumn,12pt]{IEEEtran}
\ifCLASSINFOpdf
\else
\fi

\usepackage{graphicx}%插入图片
\usepackage{bm}%粗体字包
\usepackage{amsfonts}
\usepackage{amsmath}
\usepackage{amssymb}
\usepackage{times}
\usepackage{subfigure}
\usepackage{latexsym,bm,amsmath,amssymb} %Latex 中数学包
\usepackage{CJK}
\usepackage{hhline}

\usepackage{multirow} %multirow for format of table
\usepackage{amsmath}
\usepackage{xcolor}
\usepackage{epstopdf}
\usepackage{cite}
\usepackage[noend]{algpseudocode}
\usepackage{algorithmicx,algorithm}
\usepackage{geometry}
\begin{document}
\title{Cross Domain Iterative Detection for Orthogonal Time Frequency Space Modulation}

%\author{
%Author1, Author2, Author3, Author4
%}

\author{Shuangyang~Li,~\IEEEmembership{Student Member,~IEEE,}
Weijie~Yuan,~\IEEEmembership{Member,~IEEE,}
Zhiqiang~Wei,~\IEEEmembership{Member,~IEEE,}
and
Jinhong~Yuan,~\IEEEmembership{Fellow,~IEEE}

%\thanks{S.~Li is with the School of Electrical Engineering and Telecommunications, University of New South
%Wales, Sydney, NSW 2052, Australia (e-mail: shuangyang.li@unsw.edu.au).}
%\thanks{W.~Yuan, Z.~Wei, and J. Yuan are with the School of
%Electrical Engineering and Telecommunications, the University of New South
%Wales, Australia (e-mail: {weijie.yuan,zhiqiang.wei,j.yuan}@unsw.edu.au).}
}


\maketitle

\begin{abstract}
Recently proposed orthogonal time frequency space (OTFS) modulation has been considered as a promising candidate for accommodating various emerging communication and sensing applications in high-mobility environments.
In this paper, we propose a novel cross domain iterative detection algorithm to enhance the error performance of OTFS modulation.
Different from conventional OTFS detection methods, the proposed algorithm applies basic estimation/detection approaches to both the time domain and delay-Doppler (DD) domain and iteratively updates the extrinsic information from two domains with the unitary transformation. In doing so, the proposed algorithm exploits the time domain channel sparsity and the DD domain symbol constellation constraints.
We evaluate the estimation/detection error variance in each domain for each iteration and derive the state evolution to investigate the detection error performance.
We show that the performance gain due to iterations comes from the non-Gaussian constellation constraint in the DD domain. 
More importantly, we prove the proposed algorithm can indeed converge and, in the convergence, the proposed algorithm can achieve almost the same error performance as the maximum-likelihood sequence detection even in the presence of fractional Doppler shifts.
Furthermore, the computational complexity associated with the domain transformation is low, thanks to the structure of the discrete Fourier transform (DFT) kernel.
Simulation results are consistent with our analysis and demonstrate a significant performance improvement compared to conventional OTFS detection methods.
\end{abstract}


% \fi
\IEEEpeerreviewmaketitle

\section{Introduction}
Orthogonal time frequency space (OTFS) modulation~\cite{Hadani2017orthogonal} has attracted substantial attention recently, due to its robust performance over the high-mobility environments.
In particular, beyond the fifth-generation (B5G) wireless communication systems are required to accommodate various emerging applications in high-mobility environments, such as mobile communications on board aircraft (MCA), low-earth-orbit satellites (LEOSs), high speed trains, and unmanned aerial vehicles (UAVs) %\cite{Meyer2019road,giambene2018satellite,cai2020joint}. 
\cite{Meyer2019road,cai2020joint}. In those cases, the performance of currently deployed orthogonal frequency division multiplexing (OFDM) modulation may degrade because the significant Doppler spread induced by the high-mobility can severely undermine the orthogonality between subcarriers.
Therefore, OTFS modulation has been recognized as a potential solution to supporting heterogeneous requirements of B5G wireless systems in high-mobility scenarios~\cite{Hadani2017orthogonal}.

Different from the conventional OFDM modulation, OTFS modulation considers the delay-Doppler (DD) domain signal representation instead of the time-frequency (TF) domain, where the channel responses are relatively sparse and robust~\cite{Hadani2017orthogonal}. The DD domain channel property gives rise to the symbol placement in the DD domain which allows the information symbols directly interact with the DD domain channel response, resulting in a much simpler input-output relationship compared to that of the OFDM modulation in high-mobility channels~\cite{Raviteja2019effective}.
Furthermore, by invoking the two-dimensional (2D) symplectic finite Fourier
transform (ISFFT), each DD domain symbol spreads onto the whole TF domain and thus principally experiences the whole fluctuations of the TF channel over an OTFS frame.
Therefore, OTFS modulation offers the potential of achieving the full channel diversity in a high-mobility environment~\cite{Raviteja2019effective,ShuangyangOTFS}.

Although OTFS modulation shows many advantages over the conventional OFDM modulation, it requires complex detection algorithms in order to achieve the potential full channel diversity.
With the help of cyclic prefix (CP), a single-tap frequency domain equalization is usually sufficient for OFDM modulation, where the intersymbol interference (ISI) due to the multipath effect can be largely mitigated at a cost of the signaling overhead~\cite{hwang2008ofdm}. In contrast, the DD domain channel property enables a reduced CP frame structure for OTFS modulation, where the interference due to the channel impairments has to be equalized via advanced algorithms~\cite{Zhiqiang_magzine}.
Many existing studies focused on the low-complexity detection for OTFS modulation.
%In order to reduce the detection complexity of OTFS modulation, many advanced detection algorithms has been proposed.
In \cite{Raviteja2018interference}, the authors developed an iterative receiver based on the classic message passing algorithm (MPA), where the interference from other information symbols is treated as Gaussian variables to reduce the detection complexity. However, due to the short cycles on probabilistic graphical model, MPA may fail to converge and results in performance degradation.
To solve this issue, the authors of \cite{Yuan2019simple} proposed a convergence guaranteed receiver based on the variational Bayes framework.
%The basic idea of this detector is to approximate the corresponding \emph{a posteriori} distribution of the optimal detection by exploiting the Kullback-Leibler (KL) divergence such that the MPA can be implemented based on a simpler graphical model. %compared to that of the original OTFS modulation.
An approximate message passing (AMP)-based approach was developed in~\cite{yuan2020iterative}, which can be efficiently implemented and can obtain the minimum mean square error (MMSE) detection performance with a reduced complexity.
We notice that most of the existing works on the OTFS detection take advantage of the DD domain channel sparsity to reduce the detection complexity.
However, when an OTFS frame duration is not sufficiently long, the resultant DD domain effective channel can be dense due to the insufficient resolution of the Doppler frequency, i.e., fractional Doppler~\cite{Raviteja2018interference}. In such a case, conventional detection methods may experience a very high detection complexity since the channel sparsity no longer holds.
This fact motivates us to consider the OTFS detection based on different domains. %We notice that the time domain effective channel maintains the sparsity even in the case of fractional Doppler, but the corresponding time domain OTFS signal is hard to be characterized by a discrete set due to the domain transformation. In particular,
%the transformation between the DD domain and time domain is unitary.

In this paper, we propose a novel cross domain iterative detection algorithm for OTFS modulation, where the extrinsic information is passed between the time domain and DD domain via the corresponding unitary transformations.
This special detection structure is motivated by the connection between the orthogonality and message passing based on the view from the orthogonal approximate message passing (OAMP) algorithm proposed by Ma and Li~\cite{Ma2017orthogonal}.
In particular, the rationale behind our work is that the unitary transformation between the time domain and DD domain allows the detection/estimation errors in one domain to be principally \emph{orthogonal} to the detection/estimation errors in the other domain, which can suppress the correlation between the input errors and the output errors for each domain detection/estimation~\cite{Ma2015Turbo,Ma2017orthogonal}.
Therefore, the detection/estimation instability due to the positive feedback effect, which usually caused by the correlation between the input and output errors for the iterative receivers, can be greatly reduced during iterations, and this stability in return improves the error performance~\cite{Ma2015Turbo,Ma2017orthogonal}.
In specific, we apply conventional linear minimum mean squared error (L-MMSE) estimator for equalization in the time domain, while adopt a simple \emph{symbol-by-symbol} detection algorithm in the DD domain. Interestingly, by combining such two basic methods, the proposed algorithm shows promising error performance even in very severe and complex fractional Doppler cases.
Note that conventional iterative receiver improves the error performance by iteratively exchanging the extrinsic information between two disjoint components, such as the detector and decoder, via interleaving/de-interleaving. In contrast, our proposed iterative algorithm exchanges the extrinsic information within the same component (detector) but between two orthogonal domains via unitary transformation. In other words, we separate the OTFS detection problem into two parts, corresponding to the time domain (performing de-correlation) and the DD domain (performing de-noising) and iteratively exchange the extrinsic information via unitary transformation.
In particular, we provide a detailed proof that explains the advantage of the proposed algorithm and briefly discuss its detection complexity.
The main contributions of this paper are summarized as follows.
\begin{itemize}
\item Based on the time domain sparsity, we propose a cross domain iterative detection algorithm that works in both the time domain and DD domain. Furthermore, according to the unitary property of the domain transformation, the details of cross domain message passing are discussed. We also prove that the symbol-by-symbol DD domain detection cannot provide any extrinsic information itself and therefore the extrinsic information need to be calculated in the time domain.
\item We provide theoretical analysis for the MMSE performance of the proposed algorithm based on the state evolution~\cite{Ma2015Turbo}. In particular, we derive the average MMSE per iteration and prove that the proposed algorithm can converge after a few iterations. Furthermore, we show that the average MMSE for both the time domain and the DD domain share the same value in the convergence and the error performance improvement brought by the cross domain message passing is originated from the non-Gaussian distribution of practical DD domain signal constellations.
\item We investigate the effective DD domain signal-to-noise ratio (SNR) in order to study the error performance in the convergence. We prove that the corresponding effective SNR can approach the maximum receiver SNR for a given fading channel, i.e., almost all the energy from different paths are collected and coherently combined, which indicates that the proposed algorithm can theoretically approach the error performance of the optimal maximum-likelihood sequence estimation (MLSE) detection.
    On the other hand, we show that the computational complexity of the proposed algorithm is much lower compared to that of the MLSE detection thanks to the efficient implementation based on the discrete Fourier transform (DFT) kernel for cross domain message passing between the time domain and DD domain.
    We also show that the overall detection complexity of the proposed algorithm does not increase in the presence of fractional Doppler.
\item The error performance of the proposed algorithm is evaluated by numerical simulations. Simulation results agree with our theoretical analysis and demonstrate a significant performance improvement compared to conventional OTFS detection methods.
\end{itemize}

The rest of this paper is organized as follows. We provide a brief overview and the system model of OTFS modulation in Section II.
In Section III, the proposed cross domain iterative detection algorithm is presented.
The performance analysis of the proposed algorithm is given in Section IV and finally a summary is provided in Section V.

\emph{Notations:} The blackboard bold letter ${\mathbb{A}}$ denotes an energy normalized constellation set, whose size is ${{\cal X}\left( {\mathbb{A}} \right)}$;
The blackboard bold letter ${\mathbb{C}}$ denotes the complex number field;
Boldface capitals and lower-case letters are used to define a matrix and a vector, respectively;
The blackboard bold letter ${\mathbb E}\left[ {\cdot} \right]$ denotes the expectation operator;
The notations $(\cdot)^{\rm{T}}$, $(\cdot)^{*}$, $\left\| {\cdot} \right\|$, $(\cdot)^{-1}$, and $(\cdot)^{\rm{H}}$ denote the transpose, the conjugate, the Euclidean norm, the inverse, and the Hermitian operations for a matrix, respectively; ${{{\bf{F}}_N}}$ and ${{{\bf{I}}_M}}$ denote the normalized discrete Fourier transform (DFT) matrix of size $N\times N$ and the identity matrix of size $M\times M$, respectively;
``$ \otimes $" denotes the Kronecker product operator; $\textrm{vec}(\cdot)$ denote the vectorization operation;
 $\textrm{Tr}(\cdot)$ denotes the trace operation; $\Pr(\cdot)$ denotes the probability of an event;
 $\propto$ represents both sides of the equation are multiplicatively connected to a constant;
$f(x)$ denotes an function of $x$ and its second-order derivative with respect to $x$ is denoted by ${f^{''}}\left( x  \right)$;
The big-O notation ${\cal O}\left( \cdot \right)$ asymptotically describes the order of computational complexity.
%$\mathbb{E}[\cdot]$ denote the expectation; $\textrm{det}(\cdot)$, $\textrm{tr}(\cdot)$, and $\textrm{vec}(\cdot)$ denote the determinant, the trace, and the vectorization operation; $\textrm{diag}{\{\cdot\}}$ denotes the diagonal matrix; ``$ \otimes $" denotes the Kronecker product operator; $Q(\cdot)$ denotes the tail distribution function of the standard normal distribution, $\delta(\cdot)$ denotes the Dirac delta function, ${I_0}\left( {\cdot} \right)$ denotes the zero-order modified Bessel function of the first kind, respectively; $P(\cdot)$ denotes the probability and $p(\cdot)$ denotes the probability density function (PDF); $f(\cdot)$ denotes an arbitrary function; ${\left( {f(\cdot)} \right)_{\max }}$ and ${\left( {f(\cdot)} \right)_{\min }}$ denote the maximum and minimum values of function $f(\cdot)$, respectively.
\section{System Model}
\begin{figure}
\centering
\includegraphics[width=0.8\textwidth]{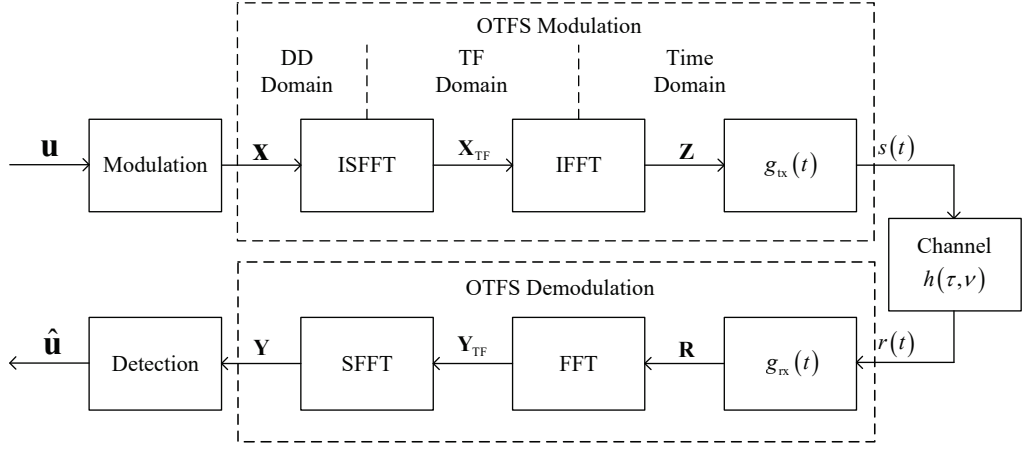}
\caption{An OTFS system model.}
\label{OTFS_diagram}
\centering
\end{figure}

Without loss of generality, the considered OTFS system diagram is given in Fig. \ref{OTFS_diagram}.
Let $N$ be the number of time slots and $M$ be the number of sub-carriers for each OTFS frame, respectively.
Let $\bf{u}$ be the information bit sequence that is modulated into the DD domain transmitted symbol vector $\bf x$, where $\bf x$ is of length $MN$, i.e., ${\bf{x}} \in {{\mathbb A}^{MN}}$. In specific, the DD domain transmitted symbol vector $\bf x$ can be arranged as a two-dimensional (2D) matrix ${\bf{X}} \in {{\mathbb A}^{M \times N}}$, i.e.,
${\bf{x}} \buildrel \Delta \over = {\rm{vec}}\left( {\bf{X}} \right) $ and the $(k,l)$-th element $x\left[ {k,l} \right]$ of ${\bf{X}}$ is the DD domain transmitted symbol in the $k$-th Doppler and $l$-th delay grid \cite{Hadani2017orthogonal}, for $0 \le k \le N-1$, and $0 \le l \le M-1$.
Then, we can obtain the TF domain transmitted symbol $X\left[ {n,m} \right], 0 \le n \le N-1$, and $0 \le m \le M-1$, according to ${\bf{X}}$ via the ISFFT \cite{Hadani2017orthogonal}, i.e.,
\begin{equation}
X\left[ {n,m} \right] = \frac{1}{{\sqrt {NM} }}\sum\limits_{k = 0}^{N - 1} {\sum\limits_{l = 0}^{M - 1} {x\left[ {k,l} \right]} } {e^{j2\pi \left( {\frac{{nk}}{N} - \frac{{ml}}{M}} \right)}}  .
\label{transmitted_symbols_TF}
\end{equation}
With the TF domain transmitted symbols, the OTFS signal $s(t)$ can be produced by the conventional OFDM modulator, such as
\begin{equation}
s\left( t \right) = \sum\limits_{n = 0}^{N - 1} {\sum\limits_{m = 0}^{M - 1} {X\left[ {n,m} \right]{g_{{\rm{tx}}}}\left( {t - nT} \right){e^{j2\pi m\Delta f\left( {t - nT} \right)}}} },
\end{equation}
where ${\Delta f}$ is the frequency spacing between adjacent sub-carriers and $g_{{\rm{tx}}}(t)$ is the transmitter shaping pulse.

We consider the OTFS signal transmitting over a time-varying channel, whose response is fully characterized by its DD domain representation \cite{Hadani2017orthogonal}, i.e.,
\begin{equation}
h\left( {\tau ,\nu } \right) = \sum\limits_{i = 1}^P {{h_i}\delta \left( {\tau  - {\tau _i}} \right)\delta \left( {\nu  - {\nu _i}} \right)}.
\label{channel}
\end{equation}
In (\ref{channel}), $P$ is the number of paths, $h_i$, $\tau _i$, and $\nu _i$ are the path gain, delay and Doppler shift corresponding to the $i$-th path, respectively.
Specifically, we denote by $l_i$ and $k_i$ the indices of delay and Doppler, respectively, where we have
\begin{equation}
{\tau _i} = \frac{{l_i}}{{M\Delta f}},\quad {\rm and }\quad
{\nu _i} = \frac{{{k_i} + {\kappa _i}}}{{NT}}.
\label{resolution}
\end{equation}
Note that the term $- {1 \mathord{\left/
 {\vphantom {1 2}} \right.
 \kern-\nulldelimiterspace} 2} \le {\kappa _i} \le {1 \mathord{\left/
 {\vphantom {1 2}} \right.
 \kern-\nulldelimiterspace} 2}$ denotes the fractional Doppler which corresponds to the fractional shift from the nearest Doppler grid \cite{Raviteja2018interference}. On the other hand, since the typical value of the sampling time ${1 \mathord{\left/
 {\vphantom {1 {M\Delta f}}} \right.
 \kern-\nulldelimiterspace} {M\Delta f}}$ in the delay domain is usually sufficiently small, the impact of
fractional delays in typical wide-band systems can be neglected \cite{tse2005fundamentals}.
%To facilitate the following analysis, we assume that the delay and Doppler indices $l_i$ and $k_i$ follow the discrete uniform distribution.
%In specific, we have $l_i \in \left[ {0,{l_{\max }}} \right]$, where $l_{\max }$ is the maximum delay index.
%We also have $k_i \in \left[ { - {k_{\max }},{k_{\max }}} \right]$, where $k_{\max }$ is the maximum Doppler index\footnote{The maximum Doppler shift is given by ${\nu_{\max }} = \frac{v}{c}{f_c}$, where $v$ is the relative user equipment (UE) speed, $c$ is the speed of light, and $f_c$ is the carrier frequency. Thus, the maximum Doppler index is given by ${k_{\max }} = {\nu_{\max }} N T$ \cite{Raviteja2018interference}. The maximum delay shift is given by ${\tau_{\max }} = \frac{d_{\rm{max}}}{c}$, where $d_{\rm{max}}$ is the maximum propagation distance difference among the $P$ channel paths.}.

Let us turn our attention to the receiver side.
Let $w\left( t \right)$ be the additive white Gaussian noise (AWGN) process with one-sided power spectral density (PSD) $N_0$. The received signal can then be written as
\begin{equation}
r\left( t \right) = \int {\int {h\left( {\tau ,\nu } \right)s\left( {t - \tau } \right)} } {e^{j2\pi \nu \left( {t - \tau } \right)}}d\tau d\nu + w\left( t \right).
\label{received_signal}
\end{equation}
Let $g_{{\rm{rx}}}(t)$ be the filter adopted at the receiver side. The received symbols $Y\left[ {n,m} \right]$ in the TF domain are then obtained by
\begin{align}
Y\left[ {n,m} \right] = \int {r\left( t \right)g_{{\rm{rx}}}^*\left( {t - nT} \right){e^{ - j2\pi m\Delta f\left( {t - nT} \right)}}} {\rm d}t .
\label{receivded_symbols_TF}
\end{align}
By performing SFFT to $Y\left[ {n,m} \right]$, we can obtain the DD domain received symbols as
\begin{equation}
y\left[ {k,l} \right] = \frac{1}{{\sqrt {NM} }}\sum\limits_{n = 0}^{N - 1} {\sum\limits_{m = 0}^{M - 1} {Y\left[ {n,m} \right]{e^{ - j2\pi \left( {\frac{{nk}}{N} - \frac{{ml}}{M}} \right)}}} } + \tilde w\left[ {k,l} \right],
\label{DD_y}
\end{equation}
where $\tilde  w\left[ {k,l} \right]$ denotes the equivalent AWGN samples in the DD domain.
%In specific, the DD domain received symbols $y\left[ {k,l} \right]$ can be arranged into the 2D received symbol matrix ${\bf{Y}}$ according to the DD grid,
%whose $(k,l)$-th element is $y\left[ {k,l} \right]$.

For the ease of derivation, we are interested in the vector form representation of the input-output relationship of the OTFS system in the corresponding domains.
We use ${\bf X}\in {{\mathbb A}^{M \times N}}$ and ${\bf Y}\in {{\mathbb C}^{M \times N}}$ to denote the DD domain transmitted symbol matrix and received symbol matrix, respectively.
Furthermore, we use the ${\bf X}_{\rm TF}\in {{\mathbb C}^{M \times N}}$ and ${\bf Y}_{\rm TF}\in {{\mathbb C}^{M \times N}}$ to denote the TF domain transmitted symbol matrix and received symbol matrix, respectively.
Similarly, the time domain transmitted symbol matrix and received symbol matrix are respectively denoted by ${\bf Z}\in {{\mathbb C}^{M \times N}}$ and ${\bf R}\in {{\mathbb C}^{M \times N}}$.
According to~\cite{Raviteja2019practical}, two normalized DFT matrices ${\bf{F}}_M$ and ${\bf{F}}_N$ of size $M \times M$ and $N \times N$ can be used to characterize the SFFT in~\eqref{transmitted_symbols_TF}. Thus, we have
\begin{equation}
{{\bf{X}}_{{\rm{TF}}}} = {{\bf{F}}_M}{\bf{XF}}_N^{\rm{H}}.
\label{TF_symbol_matrix}
\end{equation}
Then, by considering the rectangular pulse for the transmitter shaping pulse, the time domain transmitted symbol matrix can be obtained by~\cite{Raviteja2019practical}
\begin{equation}
{\bf{Z}} = {{\bf{I}}_M}{\bf{F}}_M^{\rm{H}}{{\bf{X}}_{{\rm{TF}}}} = {\bf{XF}}_N^{\rm{H}}.
\label{TD_symbol_matrix}
\end{equation}
According to~\eqref{TF_symbol_matrix} and~\eqref{TD_symbol_matrix}, we can obtain the corresponding vector form of the transmitted symbols in the DD, TF, and time domains respectively, i.e.,
\begin{align}
{\bf{x}} &\buildrel \Delta \over = {\rm{vec}}\left( {\bf{X}} \right), \label{DD_transmitted_symbol_vec}\\
{{\bf{x}}_{{\rm{TF}}}} &\buildrel \Delta \over = {\rm{vec}}\left( {{{\bf{X}}_{{\rm{TF}}}}} \right) = \left( {{\bf{F}}_N^{\rm{H}} \otimes {{\bf{F}}_M}} \right){\bf{x}}, \; {\rm and } \quad\label{TF_transmitted_symbol_vec}\\
{\bf{z}} &\buildrel \Delta \over = {\rm{vec}}\left( {\bf{Z}} \right) = \left( {{\bf{F}}_N^{\rm{H}} \otimes {{\bf{I}}_M}} \right){\bf{x}}, \label{T_transmitted_symbol_vec}
\end{align}
respectively. In particular, corresponding to the time domain transmitted symbol vector $\bf z$, the time domain effective channel ${\bf{H}}_{\rm{T}}^{{\rm{eff}}}$ with a reduced CP frame format can be given by~\cite{Raviteja2019practical}
\begin{equation}
{\bf{H}}_{\rm{T}}^{{\rm{eff}}} = \sum\limits_{i = 1}^P {{h_i}} {{\bm{\Pi }}^{{l_i}}}{{\bm{\Delta}} ^{{k_i}+{\kappa _i}}}, \label{Time_domain_channel}
\end{equation}
where ${\bm{\Pi }}$ is the permutation matrix (forward cyclic shift), i.e.,
\begin{equation}
{\bm{\Pi }} = {\left[ {\begin{array}{*{20}{c}}
0& \cdots &0&1\\
1& \ddots &0&0\\
 \vdots & \ddots & \ddots & \vdots \\
0& \cdots &1&0
\end{array}} \right]_{MN \times MN}},
\end{equation}
and ${\bm{\Delta}}=\textrm{diag}\{{\alpha}^0,{\alpha}^1,...,{\alpha}^{MN-1}\} $ is a diagonal matrix with ${\alpha} \buildrel \Delta \over = {e^{\frac{{j2\pi }}{{MN}}}}$~\cite{Raviteja2019practical}.
Thus, the received time domain symbol vector $\bf r$ is given by
\begin{equation}
{\bf{r}} = {\bf{H}}_{\rm{T}}^{{\rm{eff}}}{\bf{z}} + {\bf{w}} \label{TD_input_output_relationship},
\end{equation}
where ${\bf{w}}$ is the corresponding AWGN sample vectors in the time domain.
Specifically, $\bf r$ can be rearranged into the 2D time domain received symbol matrix $\bf R$.
By applying the rectangular pulse as the receiver filtering pulse, we can obtain the corresponding TF domain and DD domain received symbol matrices as
\begin{align}
{{\bf{Y}}_{{\rm{TF}}}} &= {{\bf{F}}_M}{{\bf{I}}_M}{\bf{R}} = {{\bf{F}}_M}{\bf{R}}, \notag\\
{\bf{Y}} &= {\bf{F}}_M^{\rm{H}}{{\bf{Y}}_{{\rm{TF}}}}{{\bf{F}}_N} = {\bf{R}}{{\bf{F}}_N}.
\end{align}
Therefore, we can derive the corresponding vector form of the received symbols in the TF domain and DD domain by
 \begin{align}
{{\bf{y}}_{{\rm{TF}}}} &\buildrel \Delta \over = {\rm{vec}}\left( {{{\bf{Y}}_{{\rm{TF}}}}} \right) = \left( {{{\bf{I}}_N} \otimes {{\bf{F}}_M}} \right){\bf{r}},\label{TF_received_symbol_vec}\\
{\bf{y}} &\buildrel \Delta \over = {\rm{vec}}\left( {\bf{Y}} \right) = \left( {{{\bf{F}}_N} \otimes {{\bf{I}}_M}} \right){\bf{r}}. \label{DD_received_symbol_vec}
\end{align}
Based on the previous analysis, we are ready to demonstrate the input-output relationship of OTFS modulation with respect to different domains.
Let us denote the effective channel matrices in TF domain and DD domain by ${\bf{H}}_{\rm{TF}}^{{\rm{eff}}} $ and ${\bf{H}}_{\rm{DD}}^{{\rm{eff}}}$, respectively.
In specific, based on~\eqref{TF_transmitted_symbol_vec},~\eqref{Time_domain_channel}, and~\eqref{TF_received_symbol_vec}, we have
\begin{align}
{\bf{H}}_{{\rm{TF}}}^{{\rm{eff}}} = \sum\limits_{i = 1}^P {{h_i}} \left( {{{\bf{I}}_N} \otimes {{\bf{F}}_M}} \right){{\bf{\Pi }}^{{l_i}}}{{\bf{\Delta }}^{{k_i} + {\kappa _i}}}\left( {{{\bf{I}}_N} \otimes {\bf{F}}_M^{\rm{H}}} \right).
\label{TF_channel_fractional}
\end{align}
Similarly, based on~\eqref{DD_transmitted_symbol_vec},~\eqref{Time_domain_channel}, and~\eqref{DD_received_symbol_vec}, we have
\begin{align}
{\bf{H}}_{\rm{DD}}^{{\rm{eff}}} = \sum\limits_{i = 1}^P {{h_i}\left( {{{\bf{F}}_N} \otimes {{\bf{I}}_M}} \right)} {{\bf{\Pi }}^{{l_i}}}{{\bf{\Delta }}^{{k_i} + {\kappa _i}}}\left( {{\bf{F}}_N^{\rm{H}} \otimes {{\bf{I}}_M}} \right).
\label{DD_channel_fractional}
\end{align}

\begin{figure}[htbp]
\centering
\subfigure[DD domain effective channel matrix ${\bf{H}}_{{\rm{DD}}}^{{\rm{eff}}}$.]{
\begin{minipage}[t]{0.5\textwidth}
\centering
\includegraphics[scale=0.5]{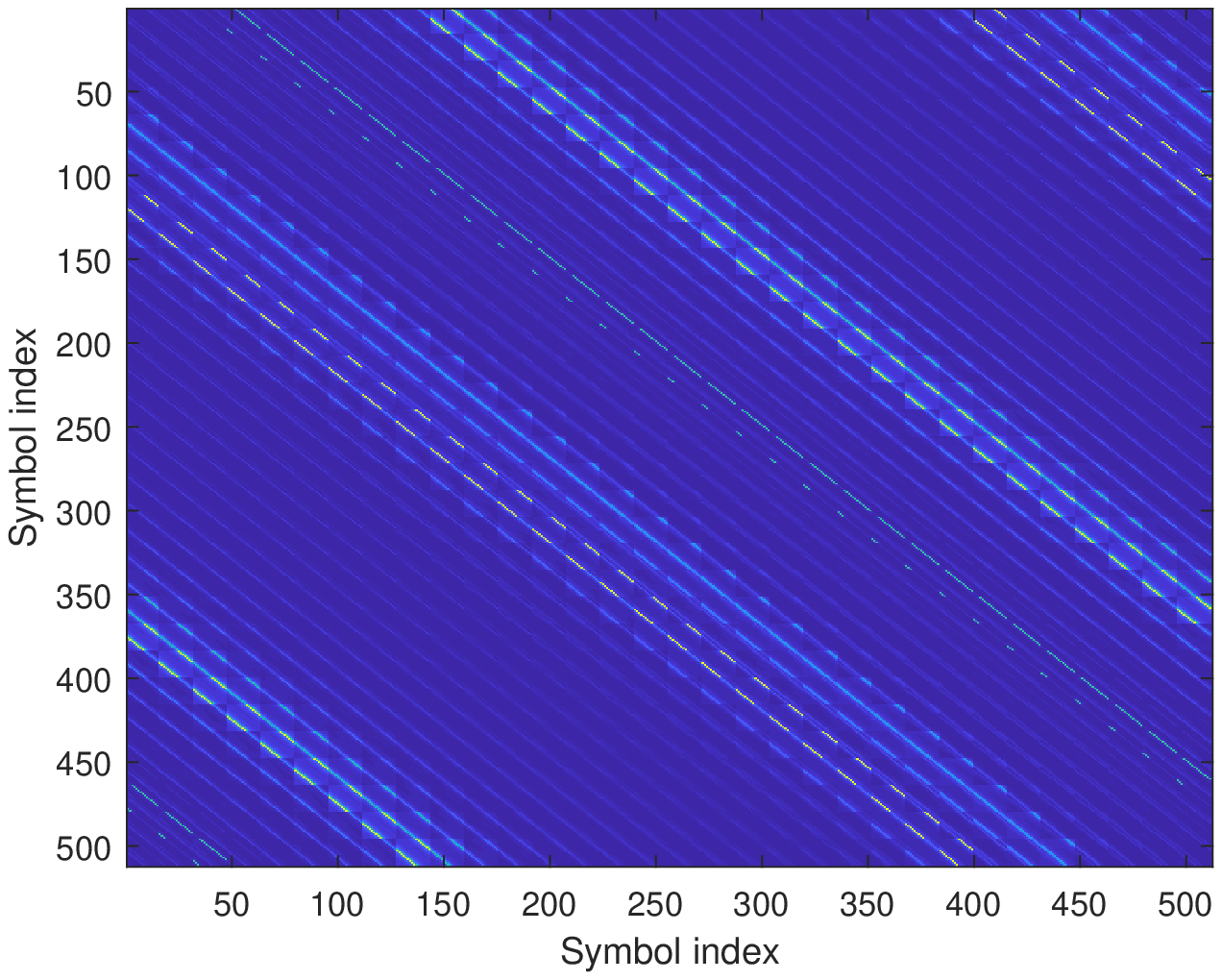}
%\caption{fig1}
\end{minipage}%
}%
\subfigure[Time domain effective channel matrix ${\bf{H}}_{{\rm{T}}}^{{\rm{eff}}}$.]{
\begin{minipage}[t]{0.5\textwidth}
\centering
\includegraphics[scale=0.5]{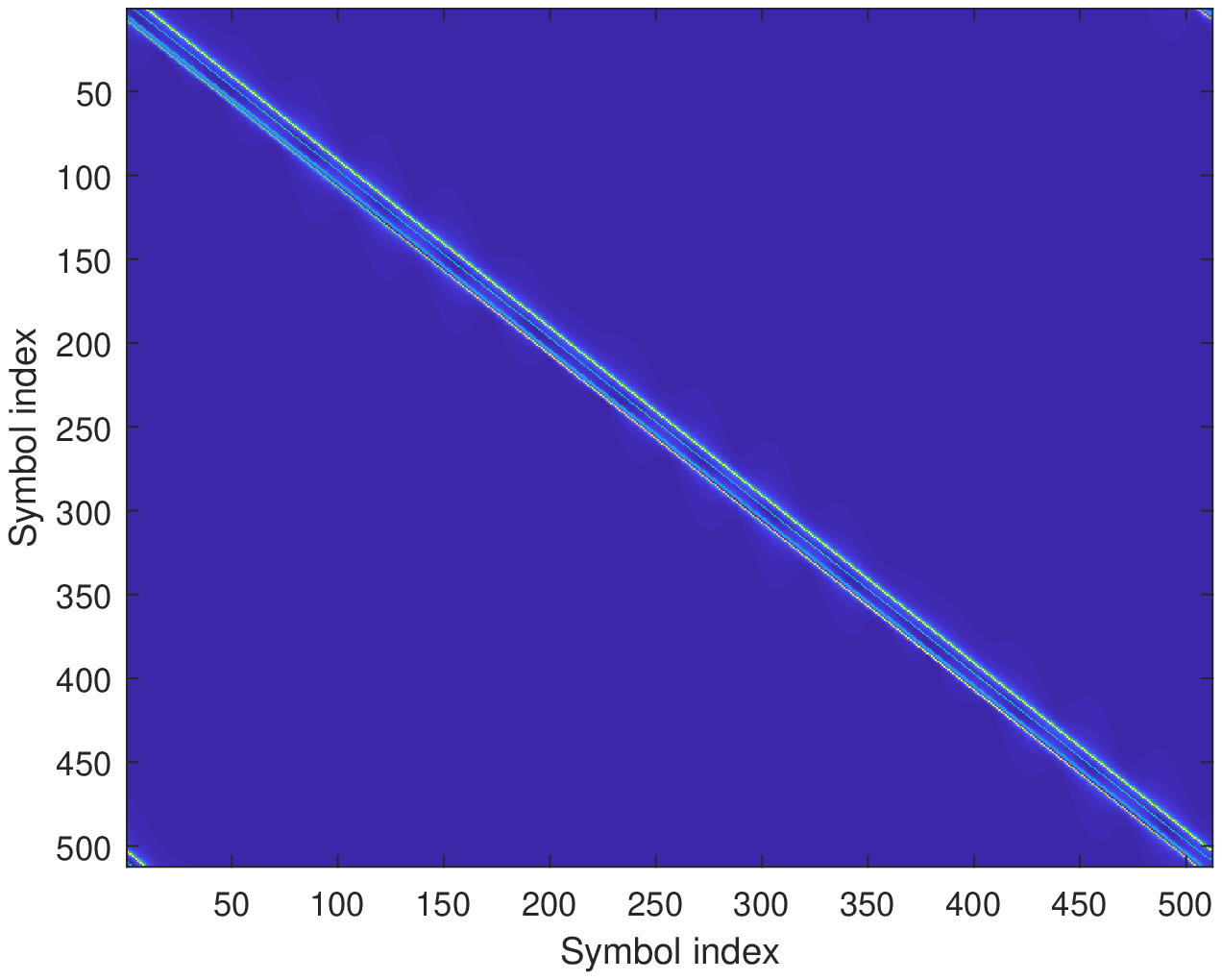}
%\caption{fig2}
\end{minipage}%
}%
\centering
\caption{Equivalent channel matrices for DD domain and time domain, where $P=5$ and the channel coefficients, delay indices and Doppler indices are $[0.28+0.45i, -0.36+0.10i, -0.34-0.23i, -0.27-0.43i, -0.93-0.54i]$, $[0,1,14,15,11]$, and $[7.33,6.92,-3.07,4.42,-9.35]$, respectively. The lines in the figures indicate the magnitudes of the corresponding matrix elements.}
\label{effective_channels}
\end{figure}

It can be shown that with the fractional Doppler shifts, the TF domain and DD domain effective channel matrices ${\bf{H}}_{{\rm{TF}}}^{{\rm{eff}}}$ and ${\bf{H}}_{{\rm{DD}}}^{{\rm{eff}}}$ as in~\eqref{TF_channel_fractional} and~\eqref{DD_channel_fractional} can be very dense. However, the time domain effective channel matrix ${\bf{H}}_{{\rm{T}}}^{{\rm{eff}}}$ in~\eqref{Time_domain_channel} remains to be sparse. Specifically, there are only at most $P$ non-zero entries in each row and column of ${\bf{H}}_{{\rm{T}}}^{{\rm{eff}}}$. For a better illustration, we show an example of ${\bf{H}}_{{\rm{DD}}}^{{\rm{eff}}}$ and ${\bf{H}}_{{\rm{T}}}^{{\rm{eff}}}$ in the presence of the fractional Doppler in Fig.~\ref{effective_channels}, where the properties of ${\bf{H}}_{{\rm{DD}}}^{{\rm{eff}}}$ and ${\bf{H}}_{{\rm{T}}}^{{\rm{eff}}}$ are clearly illustrated.
Based on the properties of the effective channel matrices, we are motivated to consider the detection based on the effective time domain channel ${\bf{H}}_{\rm{T}}^{{\rm{eff}}} $ as given by~\eqref{TD_input_output_relationship}, instead of ${\bf{H}}_{{\rm{TF}}}^{{\rm{eff}}}$ and ${\bf{H}}_{{\rm{DD}}}^{{\rm{eff}}}$ in the presence of fractional
 Doppler{\footnote {For reference, the effect of fractional Doppler can be largely mitigated by the window design~\cite{wei2020transmitter}.}}.
For notational brevity, we henceforth use $z[k]$ to denote the $k$-th entry in $\bf z$, where $0 \le k \le MN-1$.
Our proposed detection method will be presented in next section.

\section{Cross Domain Iterative Detection for OTFS Modulation}

\begin{figure*}
\centering
\includegraphics[width=0.8\textwidth]{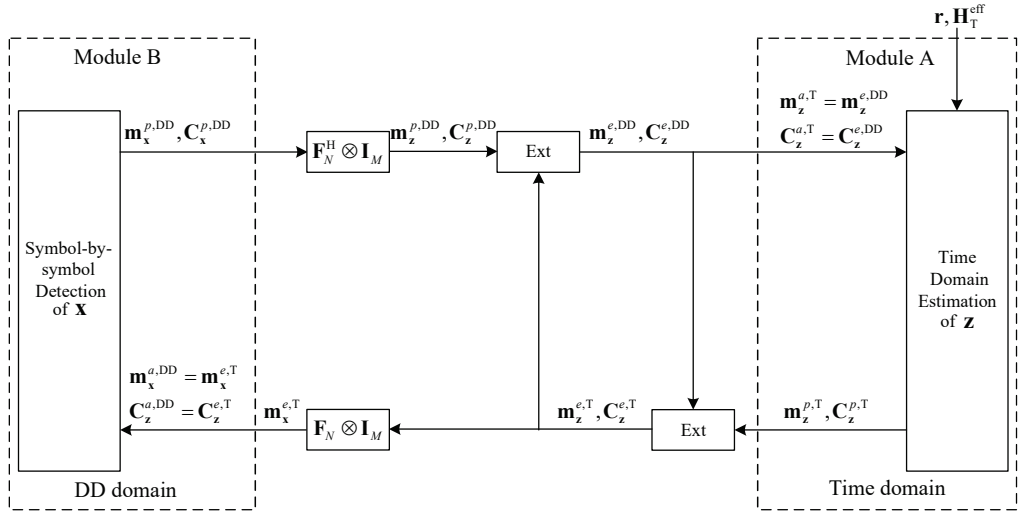}
\caption{The diagram of the proposed OTFS detector.}
\label{Detection_diagram}
\vspace{-4mm}
\centering
\end{figure*}
Notice that information symbols are multiplexed on the DD domain while the channel sparsity holds in the time domain. Therefore, we propose an iterative detector which performs de-correlation in the time domain to eliminate the effects of fading, multi-path, and Doppler, while performing de-noising in the DD domain to reduce the inaccuracy due to the noise.
%A brief diagram of the proposed cross domain iterative detection for OTFS modulation is given in Fig.~\ref{Detection_diagram}.
In particular, we assume that the entries in $\bf{x}$ independently take values from a normalized constellation set ${\mathbb{A}}$ with equal probabilities, and thus we have ${\mathbb E}\left[ {{\bf{x}}{{\bf{x}}^{\rm{H}}}} \right] = {{\bf{I}}_{MN}}$.
Then, it can be shown that entries in the time domain OTFS signal vector $\bf{z}$ are also independent from each other, i.e.,
\begin{align}
{\mathbb E}\left[ {{\bf{z}}{{\bf{z}}^{\rm{H}}}} \right] = \left( {{\bf{F}}_N^{\rm{H}} \otimes {{\bf{I}}_M}} \right){\mathbb E}\left[ {{\bf{x}}{{\bf{x}}^{\rm{H}}}} \right]\left( {{{\bf{F}}_N} \otimes {{\bf{I}}_M}} \right)={{\bf{I}}_{MN}}.\label{E_z_z}
\end{align}
Notice that the time domain OTFS signal usually behaves like a Gaussian variable due to the spreading effect of ISFFT. Therefore, we assume that the entries in $\bf{z}$ are independent and identically distributed~(i.i.d.) Gaussian variables with a unit variance according to~\eqref{E_z_z}.
We consider a detection structure that consists of two individual modules corresponding to the time domain and the DD domain, namely, module A and module B, as shown in Fig.~\ref{Detection_diagram}.
In specific, module A aims to estimate the time domain OTFS signals $\bf{z}$ and passes the estimates to module B for the detection of DD domain OTFS symbols $\bf{x}$.
On the other hand, module B carries out a simple symbol-by-symbol detection for the DD domain transmitted symbol vector $\bf{x}$ according to the estimates of time domain OTFS signal $\bf{z}$.
In Fig.~\ref{Detection_diagram}, ``Ext" denotes the calculation of extrinsic information, while ``${{\bf{F}}_N \otimes {{\bf{I}}_M}}$" and ``${{\bf{F}}_N^{\rm{H}} \otimes {{\bf{I}}_M}}$" denote the corresponding unitary transformation from the time domain to the DD domain and from the DD domain to the time domain, respectively.
For notational brevity, we use several notations denoting the \emph{a priori}, \emph{a posteriori}, and extrinsic information of the means (denoted by $\bf m$) and covariance matrices (denoted by $\bf C$) for the time domain OTFS signal vector $\bf{z}$ and DD domain symbol vector $\bf{x}$, respectively, as shown in Table~\ref{Detection_parameters}.
\begin{table*}[htbp]
\centering
\vspace{-2mm}
\caption{Notations for Proposed Algorithm Parameters}
\begin{tabular}{|l|l|l|l|l|l|l|}
\hline
  & \multicolumn{3}{c|}{Time Domain}           & \multicolumn{3}{c|}{DD Domain}             \\ \cline{2-7}
  & \emph{a priori} & \emph{a posteriori} & extrinsic (from) & \emph{a priori} & \emph{a posteriori} & extrinsic (from) \\ \hline
$\bf z$ & ${\bf{m}}_{\bf{z}}^{a,{\rm{T}}},{\bf{C}}_{\bf{z}}^{a,{\rm{T}}}$& ${\bf{m}}_{\bf{z}}^{p,{\rm{T}}},{\bf{C}}_{\bf{z}}^{p,{\rm{T}}}$&${\bf{m}}_{\bf{z}}^{e,{\rm{T}}},{\bf{C}}_{\bf{z}}^{e,{\rm{T}}}$&${\bf{m}}_{\bf{z}}^{a,{\rm{DD}}},{\bf{C}}_{\bf{z}}^{a,{\rm{DD}}}$& ${\bf{m}}_{\bf{z}}^{p,{\rm{DD}}},{\bf{C}}_{\bf{z}}^{p,{\rm{DD}}}$&${\bf{m}}_{\bf{z}}^{e,{\rm{DD}}},{\bf{C}}_{\bf{z}}^{e,{\rm{DD}}}$\\ \hline
$\bf x$ & ${\bf{m}}_{\bf{x}}^{a,{\rm{T}}},{\bf{C}}_{\bf{x}}^{a,{\rm{T}}}$& ${\bf{m}}_{\bf{x}}^{p,{\rm{T}}},{\bf{C}}_{\bf{x}}^{p,{\rm{T}}}$&${\bf{m}}_{\bf{x}}^{e,{\rm{T}}},{\bf{C}}_{\bf{x}}^{e,{\rm{T}}}$&${\bf{m}}_{\bf{x}}^{a,{\rm{DD}}},{\bf{C}}_{\bf{x}}^{a,{\rm{DD}}}$& ${\bf{m}}_{\bf{x}}^{p,{\rm{DD}}},{\bf{C}}_{\bf{x}}^{p,{\rm{DD}}}$&${\bf{m}}_{\bf{x}}^{e,{\rm{DD}}},{\bf{C}}_{\bf{x}}^{e,{\rm{DD}}}$\\ \hline
\end{tabular}
\label{Detection_parameters}
\end{table*}
The details of the two modules and the cross domain message passing will be made explicit in the coming subsections.

\subsection{Module A: L-MMSE Estimator for Time Domain OTFS Signal}
To estimate the time domain OTFS signal $\bf{z}$, we apply the conventional L-MMSE estimator in module A.
In specific, the estimation is based on the time domain received symbol vector $\bf r$ and the time domain
effective channel ${{\bf{H}}_{\rm{T}}^{{\rm{eff}}}}$ with the aid of the \emph{a priori} information, ${\bf{m}}_{\bf{z}}^{a,{\rm{T}}}$ and ${\bf{C}}_{\bf{z}}^{a,{\rm{T}}}$, that is fed back from module B.
It should be noted that ${\bf{C}}_{\bf{z}}^{a,{\rm{T}}}$ is a diagonal matrix due to the i.i.d. assumption and it is initialized as ${\bf I}_{MN}$ for the first iteration.
Therefore, it is straightforward to obtain the L-MMSE estimation matrix ${{\bf{W}}_{{\rm{MMSE}}}}$ as~\cite{kay1993fundamentals}
\begin{align}
{{\bf{W}}_{{\rm{MMSE}}}} = {\bf{C}}_{\bf{z}}^{a,{\rm{T}}}{\left( {{\bf{H}}_{\rm{T}}^{{\rm{eff}}}} \right)^{\rm{H}}}{\left( {{\bf{H}}_{\rm{T}}^{{\rm{eff}}}{\bf{C}}_{\bf{z}}^{a,{\rm{T}}}{{\left( {{\bf{H}}_{\rm{T}}^{{\rm{eff}}}} \right)}^{\rm{H}}} + {N_0}{{\bf{I}}_{MN}}} \right)^{ - 1}}.
\label{MMSE_W}
\end{align}
Furthermore, it can be shown that the \emph{a posteriori} estimation output ${\bf{m}}_{\bf{z}}^{p, {\rm T}}$ of $\bf z$ is given by
\begin{align}
{\bf{m}}_{\bf{z}}^{p,{\rm{T}}} &= {\bf{m}}_{\bf{z}}^{a,{\rm{T}}} + {{\bf{W}}_{{\rm{MMSE}}}}\left( {{\bf{r}} - {\bf{H}}_{\rm{T}}^{{\rm{eff}}}{\bf{m}}_{\bf{z}}^{a,{\rm{T}}}} \right) \notag\\
& = {\bf{m}}_{\bf{z}}^{a,{\rm{T}}} + {\bf{C}}_{\bf{z}}^{a,{\rm{T}}}{\left( {{\bf{H}}_{\rm{T}}^{{\rm{eff}}}} \right)^{\rm{H}}}{\left( {{\bf{H}}_{\rm{T}}^{{\rm{eff}}}{\bf{C}}_{\bf{z}}^{a,{\rm{T}}}{{\left( {{\bf{H}}_{\rm{T}}^{{\rm{eff}}}} \right)}^{\rm{H}}} + {N_0}{{\bf{I}}_{MN}}} \right)^{ - 1}}\left( {{\bf{r}} - {\bf{H}}_{\rm{T}}^{{\rm{eff}}}{\bf{m}}_{\bf{z}}^{a,{\rm{T}}}} \right),
\label{MMSE_output_mean}
\end{align}
and the \emph{a posteriori} covariance matrix ${\bf{C}}_{\bf{z}}^{p, {\rm T}}$ associated with $\bf z$ is given by
%\begin{align}
%{\bf{C}}_{\bf{z}}^{{\rm{post,A}}}&= {{\bf{W}}_{{\rm{MMSE}}}}{\bf{H}}_{\rm{T}}^{{\rm{eff}}}{\bf{C}}_{\bf{z}}^{{\rm{pri}},{\rm{A}}}\notag\\
%&= {\bf{C}}_{\bf{z}}^{{\rm{pri}},{\rm{A}}}-{\bf{C}}_{\bf{z}}^{{\rm{pri}},{\rm{A}}}{\left( {{\bf{H}}_{\rm{T}}^{{\rm{eff}}}} \right)^{\rm{H}}}{\left( {{\bf{H}}_{\rm{T}}^{{\rm{eff}}}{\bf{C}}_{\bf{z}}^{{\rm{pri}},{\rm{A}}}{{\left( {{\bf{H}}_{\rm{T}}^{{\rm{eff}}}} \right)}^{\rm{H}}} + {N_0}{{\bf{I}}_{MN}}} \right)^{ - 1}}{\bf{H}}_{\rm{T}}^{{\rm{eff}}}{\bf{C}}_{\bf{z}}^{{\rm{pri}},{\rm{A}}}.
%\label{MMSE_output_variance}
%\end{align}
\begin{align}
{\bf{C}}_{\bf{z}}^{p, {\rm T}}= {\bf{C}}_{\bf{z}}^{a,{\rm{T}}}-{\bf{C}}_{\bf{z}}^{a,{\rm{T}}}{\left( {{\bf{H}}_{\rm{T}}^{{\rm{eff}}}} \right)^{\rm{H}}}{\left( {{\bf{H}}_{\rm{T}}^{{\rm{eff}}}{\bf{C}}_{\bf{z}}^{a,{\rm{T}}}{{\left( {{\bf{H}}_{\rm{T}}^{{\rm{eff}}}} \right)}^{\rm{H}}} + {N_0}{{\bf{I}}_{MN}}} \right)^{ - 1}}{\bf{H}}_{\rm{T}}^{{\rm{eff}}}{\bf{C}}_{\bf{z}}^{a,{\rm{T}}}.
\label{MMSE_output_variance}
\end{align}
It should be noted that the diagonal entries of ${\bf{C}}_{\bf{z}}^{p, {\rm T}}$ are the \emph{a posteriori} MSEs of the estimates of $\bf z$ after the L-MMSE estimation,
while the non-diagonal entries can be discarded (treated as zeros), because only the diagonal entries are of interest according to the i.i.d. assumption~\cite{Ma2015Turbo}.
The details of the L-MMSE estimation are summarized in Algorithm~\ref{L_MMSE_algorithm}.

\begin{algorithm}[htb]
\caption{L-MMSE Estimation for Time Domain OTFS Signal $\bf z$}
\hspace*{0.02in} {\bf Input:}
$\bf r$, ${{\bf{H}}_{\rm{T}}^{{\rm{eff}}}}$, ${\bf{m}}_{\bf{z}}^{a,{\rm{T}}}$ and ${\bf{C}}_{\bf{z}}^{a,{\rm{T}}}$\\
\hspace*{0.02in} {\bf Steps:}
\begin{algorithmic}[1]
\State Compute the L-MMSE estimator matrix ${{\bf{W}}_{{\rm{MMSE}}}}$ by~\eqref{MMSE_W}.
\State Calculate the estimation output ${\bf{m}}_{\bf{z}}^{p,{\rm{T}}}$ by~\eqref{MMSE_output_mean}.
\State Calculate the MSE matrix ${\bf{C}}_{\bf{z}}^{p,{\rm{T}}}$ by~\eqref{MMSE_output_variance}.
\State \Return ${\bf{m}}_{\bf{z}}^{p,{\rm{T}}}$ and ${\bf{C}}_{\bf{z}}^{p,{\rm{T}}}$.
\end{algorithmic}
\label{L_MMSE_algorithm}
\end{algorithm}

\subsection{Cross Domain Message Passing: from Time Domain to DD Domain}
Based on the time domain L-MMSE estimation, we can obtain the \emph{a posteriori} information of time domain OTFS signal vector $\bf z$. However, in order to perform the iterative detection, it is important to pass the extrinsic information rather than the \emph{a posteriori} information between two the modules.
Let us define ${\bf{m}}_{\bf{z}}^{e,{\rm{T}}}$ and ${{\bf{C}}_{\bf{z}}^{e,{\rm{T}}}}$ as the extrinsic mean and covariance matrix from the L-MMSE estimation. Then, we have~\cite{kay1993fundamentals}
\begin{align}
{\bf{C}}_{\bf{z}}^{e,{\rm{T}}}&= {\left( {{{\left( {{\bf{C}}_{\bf{z}}^{p,{\rm{T}}}} \right)}^{ - 1}} - {{\left( {{\bf{C}}_{\bf{z}}^{a,{\rm{T}}}} \right)}^{{\rm{ - 1}}}}} \right)^{ - 1}} ,\label{c_z_ext_A}\\
{\bf{m}}_{\bf{z}}^{e,{\rm{T}}} &= {\bf{C}}_{\bf{z}}^{e,{\rm{T}}}\left( {{{\left( {{\bf{C}}_{\bf{z}}^{p,{\rm{T}}}} \right)}^{ - 1}}{\bf{m}}_{\bf{z}}^{p,{\rm{T}}} - {{\left( {{\bf{C}}_{\bf{z}}^{a,{\rm{T}}}} \right)}^{ - 1}}{\bf{m}}_{\bf{z}}^{a,{\rm{T}}}} \right) . \label{m_z_ext_A}
\end{align}
According to our DD domain detection formulation as will be discussed in the coming subsection, we need to obtain the DD domain \emph{a priori} mean of $\bf x$, i.e., ${\bf{m}}_{\bf{x}}^{a,{\rm{DD}}}$,
and DD domain \emph{a priori} covariance matrix of $\bf z$, i.e., ${\bf{C}}_{\bf{z}}^{a,{\rm{DD}}}$.
In specific, based on the time domain extrinsic mean of $\bf z$, we can obtain the time domain extrinsic mean of $\bf x$, which will then be forwarded to module B as the \emph{a priori} information, i.e.,
\begin{align}
{\bf{m}}_{\bf{x}}^{a,{\rm{DD}}} ={\bf{m}}_{\bf{x}}^{e,{\rm{T}}}= \left( {{{\bf{F}}_N} \otimes {{\bf{I}}_M}} \right){\bf{m}}_{\bf{z}}^{e,{\rm{T}}}. \label{m_x_pri}
\end{align}
On the other hand, the extrinsic covariance matrix ${{\bf{C}}_{\bf{z}}^{e,{\rm{T}}}}$ will be directly passed to module B as \emph{a priori} information as well. 
We have ${\bf{C}}_{\bf{z}}^{a,{\rm{DD}}}={\bf{C}}_{\bf{z}}^{e,{\rm{T}}}$.
With ${\bf{m}}_{\bf{x}}^{a,{\rm{DD}}}$ and  ${\bf{C}}_{\bf{z}}^{a,{\rm{DD}}}$ in hand, the DD domain detection is now ready to perform. 
\subsection{Module B: Symbol-by-symbol Detection for DD Domain Symbols}
By considering the relationship between the time domain OTFS signals $\bf z$ and DD domain OTFS symbols $\bf x$ in~\eqref{T_transmitted_symbol_vec}, we can
formulate the DD domain detection problem by
\begin{align}
{\bf{m}}_{\bf{z}}^{e,{\rm{T}}} = \left( {{\bf{F}}_N^{\rm{H}} \otimes {{\bf{I}}_M}} \right){\bf{x}} + {\bf{\hat w}}, \label{DD_detection_from_TD}
\end{align}
where $\bf{\hat w}$ are white Gaussian noise samples with zero mean and the variance of the $k$-th entry in $\bf{\hat w}$ is ${{C}}_{\bf{z}}^{a,{\rm{DD}}}[k,k]$.
%In~\eqref{DD_detection_from_TD}, the term ${\bf{m}}_{\bf{z}}^{{\rm{ext}},{\rm{A}}}$ is the extrinsic mean of the time domain OTFS signal $\bf z$, which is derived from the module A.
The justification of~\eqref{DD_detection_from_TD} is necessary and it is discussed in Appendix~A.
According to the ML detection rule, the detection output $\bf {\hat x}$ should satisfy
\begin{align}
{\bf{\hat x}} = \arg \mathop {\max }\limits_{\bf{x}} \Pr \left( {{\bf{m}}_{\bf{z}}^{e,{\rm{T}}}\left| {\bf{x}} \right.} \right),
\label{ML_rule}
\end{align}
where
\begin{align}
\Pr \left( {{\bf{m}}_{\bf{z}}^{e,{\rm{T}}}\left| {\bf{x}} \right.} \right) \propto \exp \left( { - \frac{1}{{{\rm{Tr}}\left( {{\bf{C}}_{\bf{z}}^{a,{\rm{DD}}}} \right)}}{{\left\| {{\bf{m}}_{\bf{z}}^{e,{\rm{T}}} - \left( {{\bf{F}}_N^{\rm{H}} \otimes {{\bf{I}}_M}} \right){\bf{x}}} \right\|}^2}} \right) . \label{MAP_rule_Forney}
\end{align}
In particular, the probability factorization based on the form of~\eqref{MAP_rule_Forney} is referred to as the Forney observation model, which was firstly introduced by Forney in~\cite{forney1972maximum}. The MLSE detection complexity of the Forney observation model based on~\eqref{MAP_rule_Forney} is exponential to the number of non-zero entries in each row of the matrix ${{\bf{F}}_N^{\rm{H}} \otimes {{\bf{I}}_M}}$. However, as the DFT matrix ${\bf{F}}_N$ is a dense matrix, the detection complexity of the Forney observation model can be very high.
Therefore, we consider a different probability factorization that is equivalent to~\eqref{MAP_rule_Forney} but only requires a linear detection complexity by taking advantage of the unitary property of the matrix ${{\bf{F}}_N^{\rm{H}} \otimes {{\bf{I}}_M}}$.

By expanding the norm operation and noticing that $\left( {{{\bf{F}}_N} \otimes {{\bf{I}}_M}} \right)\left( {{\bf{F}}_N^{\rm{H}} \otimes {{\bf{I}}_M}} \right) = {{\bf{I}}_{MN}}$,~\eqref{MAP_rule_Forney} can be equivalently expressed by
\begin{align}
\Pr \left( {{\bf{m}}_{\bf{z}}^{e,{\rm{T}}}\left| {\bf{x}} \right.} \right) \propto& \exp \left( { - \frac{{\rm{1}}}{{{\rm{Tr}}\left( {{\bf{C}}_{\bf{z}}^{a,{\rm{DD}}}} \right)}}\left( {{{\left\| {{\bf{m}}_{\bf{z}}^{e,{\rm{T}}}} \right\|}^2} - 2{\mathop{\rm Re}\nolimits} \left\{ {{{\bf{x}}^{\rm{H}}}\left( {{{\bf{F}}_N} \otimes {{\bf{I}}_M}} \right){\bf{m}}_{\bf{z}}^{e,{\rm{T}}}} \right\}} \right.} \right. \notag\\
&\quad\quad\quad\quad\quad\quad\quad\quad\quad\quad\quad\quad\quad\quad\quad\quad\quad\Bigg. {\Big. { + {{\bf{x}}^{\rm{H}}}\left( {{{\bf{F}}_N} \otimes {{\bf{I}}_M}} \right)\left( {{\bf{F}}_N^{\rm{H}} \otimes {{\bf{I}}_M}} \right){\bf{x}}} \Big)} \Bigg) \notag\\
 \propto &\exp \left( {\frac{{\rm{1}}}{{{\rm{Tr}}\left( {{\bf{C}}_{\bf{z}}^{a,{\rm{DD}}}} \right)}}\left( {2{\mathop{\rm Re}\nolimits} \left\{ {{{\bf{x}}^{\rm{H}}}\left( {{{\bf{F}}_N} \otimes {{\bf{I}}_M}} \right){\bf{m}}_{\bf{z}}^{e,{\rm{T}}}} \right\} - {{\bf{x}}^{\rm{H}}}{\bf{x}}} \right)} \right)\notag\\
\propto &\exp \left( {\frac{{\rm{1}}}{{{\rm{Tr}}\left( {{\bf{C}}_{\bf{z}}^{a,{\rm{DD}}}} \right)}}\left( {2{\mathop{\rm Re}\nolimits} \left\{ {{{\bf{x}}^{\rm{H}}}{\bf{m}}_{\bf{x}}^{a,{\rm{DD}}}} \right\} - {{\bf{x}}^{\rm{H}}}{\bf{x}}} \right)} \right).
\label{MAP_rule_Ungerboeck}
\end{align}
In particular, the probability factorization in the form of~\eqref{MAP_rule_Ungerboeck} is referred to as the Ungerboeck observation model, which was firstly introduced by Ungerboeck in~\cite{ungerboeck1974adaptive}.
%Both Forney and Ungerboeck observation models have been widely used for data detection~\cite{li2017reduced,li2020code,li2020time}, and the OTFS detection based on Forney and Ungerboeck observation models has also been considered in the previous works~\cite{Raviteja2018interference,Gaudio2020on}.
Both Forney and Ungerboeck observation models have been widely used for data detection, and the OTFS detection based on Forney and Ungerboeck observation models has also been considered in the previous works~\cite{Raviteja2018interference,Gaudio2020on}.
It should be noted that, since both ${\bf{m}}_{\bf{x}}^{a,{\rm{DD}}}$ and $\bf x$ are vectors,~\eqref{MAP_rule_Ungerboeck} can be equivalently expressed in a symbol-by-symbol fashion, such as
\begin{align}
\Pr \left( {{\bf{m}}_{\bf{z}}^{e,{\rm{T}}}\left| {\bf{x}} \right.} \right) = \prod\limits_{k = 0}^{MN - 1} {\Pr \left( {{\bf{m}}_{\bf{z}}^{e,{\rm{T}}}\left| {x\left[ k \right]} \right.} \right)} ,
\end{align}
where
\begin{align}
\Pr \left( {{\bf{m}}_{\bf{z}}^{e,{\rm{T}}}\left| {x\left[ k \right]} \right.} \right) &= \Pr \left( {m_x^{a,{\rm{DD}}}\left[ k \right]\left| {x\left[ k \right]} \right.} \right) \notag\\
&\propto  {\exp \Big( {\frac{{\rm{1}}}{{C_{\bf{z}}^{a,{\rm{DD}}}\left[ k,k \right]}}\big( {2{\mathop{\rm Re}\nolimits} \left\{ {{x^*}\left[ k \right]m_{\bf{x}}^{a,{\rm{DD}}}\left[ k \right]} \right\} - {{\left| {x\left[ k \right]} \right|}^2}} \big)} \Big)} . \label{DD_domain_detection_metric}
\end{align}
It can be observed from~\eqref{DD_domain_detection_metric} that, the optimal MLSE detection for DD domain symbols can be carried out in a simple symbol-by-symbol form, and the
corresponding inputs for detection are the DD domain symbol estimates ${\bf{m}}_{\bf{x}}^{a,{\rm{DD}}}$ and the covariance matrix ${\bf{C}}_{\bf{z}}^{a,{\rm{DD}}}$ of the time domain OTFS signal $\bf z$.

With the i.i.d. assumption of $x[k]$, the \emph{a posteriori} probability $\Pr \left( {x\left[ k \right]\left| {{\bf{m}}_{\bf{z}}^{e,{\rm{T}}}} \right.} \right)$ is essentially the same as that of ${\Pr \left( {{\bf{m}}_{\bf{z}}^{e,{\rm{T}}}\left| {x\left[ k \right]} \right.} \right)}$, i.e., $\Pr \left( {x\left[ k \right]\left| {{\bf{m}}_{\bf{z}}^{e,{\rm{T}}}} \right.} \right) \propto \Pr \left( {{\bf{m}}_{\bf{z}}^{{e},{\rm{T}}}\left| {x\left[ k \right]} \right.} \right)$.
Therefore, we can obtain the \emph{a posteriori} mean ${{m}}_{\bf{x}}^{{p,\rm{DD}}}\left[ k \right]$ of $x\left[ k \right]$ by
\begin{align}
{m}_{\bf{x}}^{p,{\rm{DD}}}\left[ k \right]  = {\mathbb E}\left[ {x\left[ k \right]|{\bf{m}}_{\bf{z}}^{e,{\rm{T}}}} \right]=\sum\limits_{i = 1}^{{\cal X}\left( {\mathbb{A}} \right)} {\Pr \left( {x\left[ k \right] = {\mathbb A}\left[ i \right]\left| {{\bf{m}}_{\bf{z}}^{e,{\rm{T}}}} \right.} \right)}  \times {\mathbb{A}}\left[ i \right],\label{m_x_post_B_element}
\end{align}
where ${\mathbb{A}}\left[ i \right]$ is the $i$-th DD domain constellation point.
Meanwhile, the \emph{a posteriori} covariance matrix ${{\bf C}}_{\bf{x}}^{p, {\rm{DD}}}$ of $\bf x$ is a diagonal matrix
due to the i.i.d. assumption, whose $k$-th element in the main diagonal is the \emph{a posteriori} variance of $x[k]$ and is given by
\begin{align}
{C}_{\bf{x}}^{p,{\rm{DD}}}\left[ {k,k} \right] &={\mathbb E}\left[ {{{\left| {x\left[ k \right] - {\mathbb E}\left[ {x\left[ k \right]|{\bf{m}}_{\bf{z}}^{e,{\rm{T}}}} \right]} \right|}^2}} \right] \notag\\
&= \sum\limits_{i = 1}^{{\cal X}\left( {\mathbb{A}} \right)} {\Pr \left( {x\left[ k \right] = {\mathbb A}\left[ i \right]\left| {{\bf{m}}_{\bf{z}}^{e,{\rm{T}}}} \right.} \right)}  \times {\left| {{\mathbb{A}}\left[ i \right]} \right|^2} - {\left| {m}_{\bf{x}}^{p,{\rm{DD}}}\left[ k \right] \right|^2}. \label{C_x_post_B_element}
\end{align}
Based on the \emph{a priori} and \emph{a posteriori} information, the extrinsic information of the DD domain detection is ready to be computed. However, we notice that the
DD domain detection with~\eqref{DD_domain_detection_metric} is a component-wise operation, and therefore in principle, the DD domain detection cannot provide any extrinsic information. The following Proposition clarifies this problem.

\textbf{Proposition 1} \emph{(Extrinsic information from the DD domain detection)}:
The DD domain detection with~\eqref{DD_domain_detection_metric} is a component-wise operation and thus it cannot provide any extrinsic information.

\emph{Proof}: The proof is given in Appendix B.

According to Proposition 1, we know that directly computing the extrinsic information based on the DD domain detection is not a good choice. Therefore, we firstly convert the \emph{a posteriori} probability of the DD domain symbols $\bf x$ to the \emph{a posteriori} probability of time domain signals $\bf z$ and then compute the extrinsic information.
In specific, the \emph{a posteriori} mean ${\bf{m}}_{\bf{x}}^{p,{\rm{DD}}}$ and covariance matrix ${{\bf C}}_{\bf{x}}^{p,{\rm{DD}}}$ will be served as the
outputs of module B. The details of how to compute the extrinsic information based on the outputs of module B will be discussed in the coming subsection.
With the above discussion, we summarize the symbol-by-symbol detection for DD domain symbols in Algorithm~\ref{DD_detection_algorithm}.
\begin{algorithm}[htb]
\caption{Symbol-by-symbol Detection for DD Domain Symbols}
\hspace*{0.02in} {\bf Input:}
${\bf{m}}_{\bf{x}}^{a,{\rm{DD}}}$, ${\bf{C}}_{\bf{z}}^{a,{\rm{T}}}$, and $\mathbb A$\\
\hspace*{0.02in} {\bf Steps:}
\begin{algorithmic}[1]
\State \textbf{for} $k$ from 0 to $MN-1$ \textbf{do}
\State $\quad$ Calculate the \emph{a posteriori} probability of $x_k$ by~\eqref{DD_domain_detection_metric}.
\State $\quad$ Make hard decision on $x_k$, which is denoted as ${\hat x}[k]$.
\State $\quad$ Compute the \emph{a posteriori} mean ${m}_{\bf{x}}^{p,{\rm{DD}}}\left[ k \right]$ of $x_k$ by~\eqref{m_x_post_B_element}.
\State $\quad$ Compute the \emph{a posteriori} variance of $x_k$ by~\eqref{C_x_post_B_element}.
\State \textbf{end for}
\State Compute  ${\bf m}_{\bf{x}}^{p,{\rm{DD}}}$ of $\bf x$, based on ${m}_{\bf{x}}^{p,{\rm{DD}}}\left[ k \right]$, $0 \le k \le MN-1$.
\State Compute  ${\bf C}_{\bf{x}}^{p,{\rm{DD}}}$ of $\bf x$, based on ${C}_{\bf{x}}^{p,{\rm{DD}}}\left[ k,k \right]$, $0 \le k \le MN-1$.
\State \Return ${\bf m}_{\bf{x}}^{p,{\rm{DD}}}$, ${\bf C}_{\bf{x}}^{p,{\rm{DD}}}$, and $\bf {\hat x}$.
\end{algorithmic}
\label{DD_detection_algorithm}
\end{algorithm}

\subsection{Cross Domain Message Passing: from DD Domain to Time Domain}

Based on the description in the previous subsection, we will discuss the message passing from DD domain to time domain.
According to~\eqref{T_transmitted_symbol_vec}, ${{\bf m}}_{\bf{x}}^{p,{\rm{DD}}}$ and ${{\bf C}}_{\bf{x}}^{p,{\rm{DD}}}$ are converted to the \emph{a posteriori} mean ${\bf{m}}_{\bf{z}}^{p,{\rm{DD}}}$ and covariance matrix ${\bf{C}}_{\bf{z}}^{p,{\rm{DD}}}$ of the time domain OTFS signal $\bf{z}$ by
\begin{align}
{\bf{m}}_{\bf{z}}^{p,{\rm{DD}}} &= \left( {{\bf{F}}_N^{\rm{H}} \otimes {{\bf{I}}_M}} \right){{\bf m}}_{\bf{x}}^{p,{\rm{DD}}} ,\label{m_z_post_B}\\
{\bf{C}}_{\bf{z}}^{p,{\rm{DD}}} &= \left( {{\bf{F}}_N^{\rm{H}} \otimes {{\bf{I}}_M}} \right){{\bf C}}_{\bf{x}}^{p,{\rm{DD}}}\left( {{{\bf{F}}_N} \otimes {{\bf{I}}_M}} \right). \label{C_z_post_B}
\end{align}
Again, we note that ${\bf{C}}_{\bf{z}}^{p,{\rm{DD}}}$ can be a dense matrix if the diagonal entries of ${{\bf C}}_{\bf{x}}^{p,{\rm{DD}}}$ are not of the same value.
In this case, we can discard the non-diagonal entries (treated as zeros), according to the i.i.d. assumption~\cite{Ma2015Turbo}.
Similar to~\eqref{c_z_ext_A} and~\eqref{m_z_ext_A} , we can compute the extrinsic information of $\bf{z}$ in terms of the mean and covariance matrix, which are given by
\begin{align}
{\bf{C}}_{\bf{z}}^{e,{\rm{DD}}}&= {\left( {{{\left( {{\bf{C}}_{\bf{z}}^{p,{\rm{DD}}}} \right)}^{ - 1}} - {{\left( {\bf{C}}_{\bf{z}}^{a,{\rm{DD}}} \right)}^{{\rm{ - 1}}}}} \right)^{ - 1}} ,\label{c_z_pri_A}\\
{\bf{m}}_{\bf{z}}^{e,{\rm{DD}}} &= {\bf{C}}_{\bf{z}}^{e,{\rm{DD}}}\left( {{{\left( {{\bf{C}}_{\bf{z}}^{p,{\rm{DD}}}} \right)}^{ - 1}}{\bf{m}}_{\bf{z}}^{p,{\rm{DD}}} - {{\left( {{\bf{C}}_{\bf{z}}^{a,{\rm{DD}}}} \right)}^{ - 1}}{\bf{m}}_{\bf{z}}^{e,{\rm{T}}}} \right) . \label{m_z_pri_A}
\end{align}
Finally, the extrinsic information of $\bf{z}$ is fed back to module A as the \emph{a priori} mean and covariance matrix for the next iteration, i.e., ${\bf{m}}_{\bf{z}}^{a,{\rm{T}}} = {\bf{m}}_{\bf{z}}^{e,{\rm{DD}}}$ and ${\bf{C}}_{\bf{z}}^{a,{\rm{T}}} = {\bf{C}}_{\bf{z}}^{e,{\rm{DD}}}$.
\begin{algorithm}[htb]
\caption{Cross Domain Iterative Detection for OTFS Modulation}
\hspace*{0.02in} {\bf Input:}
$\bf r$, ${{\bf{H}}_{\rm{T}}^{{\rm{eff}}}}$, $L_{\rm max}$, and $\mathbb A$\\
\hspace*{0.02in} {\bf Initialization:}
Set ${m}_{\bf{z}}^{a,{\rm{T}}}[k]=0$, for $0 \le k \le MN-1$ and ${\bf{C}}_{\bf{z}}^{{a},{\rm{T}}} = {{\bf{I}}_{MN}}$.\\
\hspace*{0.02in} {\bf Steps:}
\begin{algorithmic}[1]
\State \textbf{for} $l$ from 1 to $L_{\rm max}$ \textbf{do}
\State $\quad$ Perform the L-MMSE estimation for time domain OTFS signal according to Algorithm~\ref{L_MMSE_algorithm}.
\State $\quad$ Compute ${\bf{C}}_{\bf{z}}^{a,{\rm{DD}}}$ based on~\eqref{c_z_ext_A} and ${\bf{m}}_{\bf{x}}^{a,{\rm{DD}}}$ based on~\eqref{m_z_ext_A}.
\State $\quad$ Perform symbol-by-symbol detection for DD domain symbols according to Algorithm~\ref{DD_detection_algorithm}.
\State $\quad$ Compute ${\bf{m}}_{\bf{z}}^{p,{\rm{DD}}}$ and ${\bf{C}}_{\bf{z}}^{p,{\rm{DD}}}$ based on~\eqref{m_z_post_B} and~\eqref{C_z_post_B}.
\State $\quad$ Compute ${\bf{C}}_{\bf{z}}^{a,{\rm{T}}}$ and ${\bf{m}}_{\bf{z}}^{a,{\rm{T}}}$ based on~\eqref{c_z_pri_A} to~\eqref{m_z_pri_A}.
\State \textbf{end for}
\State \Return $\bf {\hat x}$.
\end{algorithmic}
\label{OTFS_detection_algorithm}
\end{algorithm}

Based on the above discussion, we summarize the proposed OTFS detection in Algorithm~\ref{OTFS_detection_algorithm}, where the term $L_{\rm max}$ is referred to as the maximum number of iterations. For notational brevity, we drop the iteration index $l$ of the corresponding matrices in Algorithm~\ref{OTFS_detection_algorithm}. So far, we have introduced the proposed cross domain iterative detection. In the following section, we will investigate its error performance and the computation complexity.

\section{Performance Analysis}
In this section, we will investigate the asymptotical error performance of the proposed detection algorithm with $MN \to \infty $ and its detection complexity.
%In particular, we focus on the error performance in the asymptotical regime, where $MN \to \infty $. In this case, the following assumption holds due to the law of large numbers.
Since the proposed algorithm involves several iterations between two modules, we characterize the error performance by the recursion of two states corresponding to each module.
Meanwhile, we will also discuss the detection complexity corresponding to each module.

\subsection{MSE Performance Analysis via State Evolution}
Without loss of generality, we first investigate the MSE performance with a given time domain effective channel ${\bf{H}}_{\rm{T}}^{{\rm{eff}}}$ and the convergence behaviour of the proposed algorithm.
%Notice that the extrinsic information is computed in the time domain.
%Therefore, we can investigate the average \emph{a priori} variance of the inputs to module A and module B during each iteration.
To this end, we can investigate the average \emph{a priori} variance of the inputs to module A and module B during each iteration.
In particular, by noticing the i.i.d. assumption of both the DD domain OTFS symbols and the time domain OTFS signal,
we define the two states for module A and module B at the $l$-th iteration by
\begin{align}
v_z^{{a},{\rm{T}}}\left( l \right)\buildrel \Delta \over = &{\mathbb E}\left[ {C_{\bf{z}}^{e,{\rm{DD}}}\left[ {k,k} \right]} \right]=\mathop {\lim }\limits_{MN \to \infty }  \frac{1}{{MN}}{\rm{Tr}}\left( {{\bf{C}}_{\bf{z}}^{{e},{\rm{DD}}}} \right),\\
v_z^{{a},{\rm{DD}}}\left( l \right) \buildrel \Delta \over = &{\mathbb E}\left[ {C_{\bf{z}}^{{e},{\rm{T}}}\left[ {k,k} \right]} \right] = \mathop {\lim }\limits_{MN \to \infty } \frac{1}{{MN}}{\rm{Tr}}\left( {{\bf{C}}_{\bf{z}}^{{e},{\rm{T}}}} \right),
\end{align}
where the expectation is with respect to the index $k$. In specific, these two states can be viewed as the average MSE of the inputs to module A and module B at the $l$-th iteration, respectively.
For notational brevity, we further define the ratios between the
OTFS signal energy and the average \emph{a priori} variance of the inputs to each module, i.e., the effective SNR for the time domain and DD domain, by
\begin{align}
{\eta _{\rm{T}}}\left( l \right) \buildrel \Delta \over = &\frac{1}{v_{z}^{a,{\rm{T}}}\left( l \right)},\\
{\eta _{\rm{DD}}}\left( l \right) \buildrel \Delta \over = &\frac{1}{v_{z}^{a,{\rm{DD}}}\left( l \right)}.
\end{align}
Now, we focus on the evolution between the two states $v_z^{{a},{\rm{T}}}\left( l \right)$ and $v_z^{{a},{\rm{DD}}}\left( l \right)$. W will investigate the corresponding average variance of the inputs and outputs for each module.
In particular, we consider the following assumption.

\textbf{Assumption 1}:
For the $l$-th iteration, the main diagonal entries of \emph{a priori} covariance matrices ${\bf{C}}_{\bf{z}}^{a,{\rm{T}}}$ and ${\bf{C}}_{\bf{z}}^{a,{\rm{DD}}}$ are of the same value as ${v_{z}^{a,{\rm{T}}}\left( l \right)}$ and ${v_{z}^{a,{\rm{DD}}}\left( l \right)}$, respectively.

It should be noted that the above assumption is reasonable with a sufficiently large number of $MN$, due to the strong law of large numbers.
With this assumption, we will discuss the connection between the two states based on Algorithm~\ref{OTFS_detection_algorithm}.
In specific, we can first derive the \emph{a posteriori} covariance matrix ${\bf{C}}_{\bf{z}}^{p,{\rm{T}}}$ according to~\eqref{MMSE_output_variance} and further derive the extrinsic covariance matrix ${\bf{C}}_{\bf{z}}^{e,{\rm{T}}}$ according to~\eqref{c_z_ext_A}, such as
\begin{align}
v_z^{{a},{\rm{DD}}}\left( l \right) = \frac{1}{{\frac{1}{{v_z^{{p},{\rm{T}}}\left( l \right)}} - \frac{1}{{v_z^{{a},{\rm{T}}}\left( l \right)}}}}, \label{state_evolution_A_B}
\end{align}
where
\begin{align}
&v_z^{{p},{\rm{T}}}\left( l \right)\notag\\
=&v_z^{{a},{\rm{T}}}\left( l \right) - \frac{{{{\left( {v_z^{{a},{\rm{T}}}\left( l \right)} \right)}^2}}}{{MN}}{\rm{Tr}}\left( {{{\left( {{\bf{H}}_{\rm{T}}^{{\rm{eff}}}} \right)}^{\rm{H}}}{{\left( {v_z^{{a},{\rm{T}}}\left( l \right){\bf{H}}_{\rm{T}}^{{\rm{eff}}}{{\left( {{\bf{H}}_{\rm{T}}^{{\rm{eff}}}} \right)}^{\rm{H}}} + {N_0}{{\bf{I}}_{MN}}} \right)}^{ - 1}}{\bf{H}}_{\rm{T}}^{{\rm{eff}}}} \right).     \label{a_posteriori_MMSE}
\end{align}
The above equations demonstrate the connection between the state $v_z^{{a},{\rm{T}}}\left( l \right)$ and $v_z^{{a},{\rm{DD}}}\left( l \right)$ at the $l$-th iteration. In the following, we will consider the update of the state $v_z^{a,{\rm{T}}}\left( l +1\right)$ based on the state $v_z^{{a},{\rm{DD}}}\left( l \right)$.
Let us define the MSE of the DD domain symbol detection, given an AWGN observation with an SNR $\eta$ by~\cite{Ma2015Turbo}
\begin{align}
MSE\left( \eta  \right) = {\mathbb E}\left[ {{{\left| {x - {\mathbb E}\left[ {x|x + \xi } \right]} \right|}^2}} \right], \label{Evolution_MMSE}
\end{align}
where $x$ is an arbitrary DD domain OTFS symbol and $\xi$ is an AWGN sample with zero mean and variance $1/{\eta}$.
According to the law of large numbers, it can be shown that the average of the main diagonal entries $v_x^{{p},{\rm{DD}}}\left( l \right)$ of the \emph{a posteriori} covariance matrix ${{\bf C}_{\bf{x}}^{p,{\rm{DD}}}}$ satisfies~\cite{Ma2015Turbo}
\begin{align}
v_x^{{p},{\rm{DD}}}\left( l \right) \buildrel \Delta \over = {\mathbb E}\left[ {C_{\bf{z}}^{{p},{\rm{DD}}}\left[ {k,k} \right]} \right] = \mathop {\lim }\limits_{MN \to \infty } \frac{1}{{MN}}{\rm{Tr}}\left( {{\bf{C}}_{\bf{x}}^{{p},{\rm{DD}}}} \right) = MSE\left( {{\eta _{\rm{DD}}}\left( l \right)} \right). \label{Average_trace}
\end{align}
Since the extrinsic information is calculated in the time domain, we need to convert the DD domain \emph{a posteriori} covariance matrix ${{\bf C}_{\bf{x}}^{p,{\rm{DD}}}}$ to the time domain \emph{a posteriori} covariance matrix ${{\bf C}_{\bf{z}}^{p,{\rm{DD}}}}$, according to~\eqref{C_z_post_B}.
Denote by ${v_z^{{p},{\rm{DD}}}\left( l \right)}$ the average of the main diagonal entries of ${{\bf C}_{\bf{z}}^{p,{\rm{DD}}}}$. According to the law of large numbers, we can show that ${v_z^{{p},{\rm{DD}}}\left( l \right)}=v_x^{{p},{\rm{DD}}}\left( l \right)$, due to the unitary transformation between the DD domain and the time domain.
Thus, according to~\eqref{c_z_pri_A}, we have
\begin{align}
v_z^{{a},{\rm{T}}}\left( {l + 1} \right) = \frac{1}{{\frac{1}{{v_z^{p},{\rm{DD}}}\left( l \right)}} - \frac{1}{{v_z^{a,{\rm{DD}}}\left( l \right)}}}. \label{v_z_pri_A}
\end{align}
Based on~\eqref{v_z_pri_A}, the state evolution from the state $v_z^{{a},{\rm{DD}}}\left( l \right)$ to the state $v_z^{{a},{\rm{T}}}\left( l +1\right)$ is now explicit.

We notice the above state evolution requires the calculation of MSE~\eqref{Evolution_MMSE}. In order to calculate the MSE, we need to compute the \emph{a posteriori} mean of $x$. However, the calculation of the \emph{a posteriori} mean is in general a nonlinear function of the observation $x + \xi$ with respect to the specific constellation shape, unless $x$ is Gaussian distributed~\cite{lozano2005mercury}.
Therefore, in order to obtain some general conclusions regarding the MSE characteristics, we consider a Monte Carlo approach to calculate the MSE.
In particular, by considering a sufficiently large value of $MN$, we produce ${m}_{\bf{x}}^{a,{\rm{DD}}}$ by using the Monte Carlo approach with a given constellation set $\mathbb A$ according to~\eqref{DD_detection_from_TD},
where the variance of $\bf{\hat w}$ is set to be $1/{\eta _{\rm{DD}}}\left( l \right)$ due to the law of large numbers.
Therefore, based on the generated ${m}_{\bf{x}}^{a,{\rm{DD}}}$, we can obtain the MSE value based on~\eqref{Evolution_MMSE}.

According to the above analysis, we notice that there exists a fixed point in the state evolution, where the overall MSE performance of the proposed algorithm does not change anymore with the increase of number of iterations $l$, i.e., the algorithm is converged. In particular,
we consider the converged MSE performance and
denote the corresponding average \emph{a posteriori} variance with respect to the time domain estimates and to the DD domain detection outputs by $v_z^{p,{\rm{T}}}$ and $v_x^{p,{\rm{DD}}}$. Then, we have the following Theorem.

\textbf{Theorem 1} \emph{(Fixed point of state evolution)}:
When the algorithm is converged, the average \emph{a posteriori} variance with respect to the time domain estimates and the DD domain detection outputs share the same value, i.e.,
\begin{align}
v_z^{p,{\rm{T}}} = v_x^{p,{\rm{DD}}}.\label{Fixed_point}
\end{align}

\emph{Proof}: Note that the values of the states $v_z^{a,{\rm{T}}}\left( l \right)$ and $v_z^{{a},{\rm{DD}}}\left( l \right)$ will not change with the increase of the iteration number if the algorithm is converged. Therefore, we combine~\eqref{state_evolution_A_B} and~\eqref{v_z_pri_A} and drop the iteration index $l$, yielding
\begin{align}
\frac{{\rm{1}}}{{\frac{1}{{v_z^{p,{\rm{T}}}}} - \frac{1}{{v_z^{a,{\rm{T}}}}}}} = \frac{{\rm{1}}}{{\frac{1}{{v_z^{p,{\rm{DD}}}}} - \frac{1}{{v_z^{a,{\rm{DD}}}}}}}.
\end{align}
After some straightforward manipulations, we can obtain~\eqref{Fixed_point}.
This completes the proof of Theorem 1.$\hfill\blacksquare$

Theorem 1 indicates that the proposed algorithm can converge and in the convergence, both time domain estimation and DD domain detection can provide the same accuracy regarding the data recovery. Other than the convergence behavior of the proposed algorithm, it is also important to derive the corresponding error performance, when the proposed algorithm is converged.
According to Theorem 1, we can evaluate this in either time domain or DD domain, when the proposed algorithm is converged. This issue will be discussed in detail in the coming subsection.
In the following, we investigate a special case where the DD domain symbols are assumed to be Gaussian distributed. We note that such a case is not practically important but it can provide some interesting insights for the analysis of the proposed algorithm. In particular, we have the following Proposition.

\textbf{Proposition 2} \emph{(Detection performance with Gaussian constellation in DD domain)}:
For the case where the DD domain symbols are Gaussian distributed, the DD domain detection cannot provide any error performance improvement, i.e., $v_z^{{p},{\rm{DD}}}\left( l \right) = v_z^{{a},{\rm{DD}}}\left( l \right)$.

\emph{Proof}: With the Gaussian assumption, we have
\begin{align}
v_z^{p,{\rm{DD}}}\left( l \right)=MSE\left( \eta  \right) = {\mathbb E}\left[ {{{\left| {x - {\mathbb E}\left[ {x|x + \xi } \right]} \right|}^2}} \right] = {\mathbb E}\left[ {{{\left| \xi  \right|}^2}} \right] = \frac{1}{\eta }=v_z^{{a},{\rm{DD}}}\left( l \right).
\end{align}
The completes the proof of Proposition 2.$\hfill\blacksquare$

Proposition 2 suggests that, if the DD domain constellation is Gaussian distributed, iteratively updating the extrinsic information between the time domain and DD domain will not introduce any error performance gain. In other words, the error performance improvement is due to the non-Gaussian constellation constraint in the DD domain.
Intuitively speaking, the DD domain detection can be viewed as a de-noising operation. If the constellation is Gaussian distributed, the ML detection will give the detection output as $x + \xi$ because $ {\mathbb E}\left[ {x|x + \xi } \right]=x + \xi$ always holds, thus it cannot correct any error induced by the noise.
On the other hand, when the DD domain constellation is not Gaussian distributed, applying iterations crossing time and DD domain can potentially improve the error performance. This is because
the time domain estimation in module A assumes that the $\bf z$ is a Gaussian vector due to the spreading effect of ISFFT, which does not take advantage of the DD domain constellation constraint. Therefore, by performing DD domain symbol detection, the DD domain constellation constraint is exploited by the proposed algorithm, which can lead to a potential error performance improvement.

\begin{figure}
\centering
\includegraphics[width=0.6\textwidth]{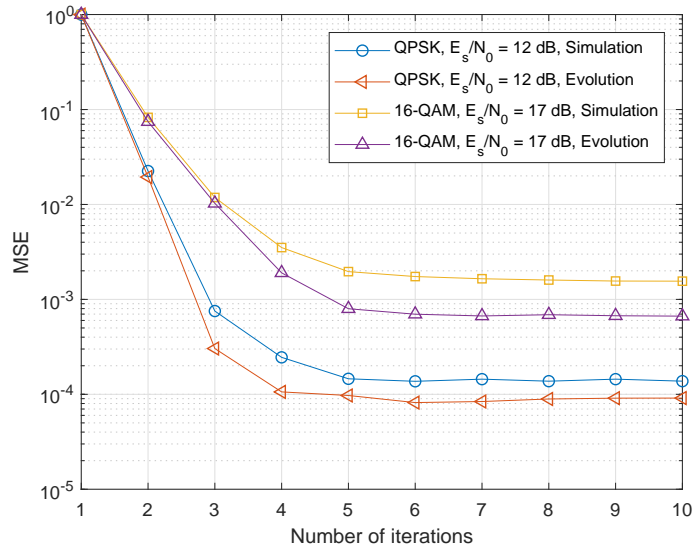}
\caption{Time domain MSE performance for OTFS modulation with $P=4$, where the frame contains $2048$ DD domain QPSK or 16-QAM symbols. In specific, the SNR for the QPSK case is
${E_s}/{N_0}=12$ dB, while that for the 16-QAM case is ${E_s}/{N_0}=17$ dB.}
\label{MSE_QPSK_16QAM}
\centering
\vspace{-3mm}
\end{figure}

Fig.~\ref{MSE_QPSK_16QAM} shows the time domain MSE performance of the proposed algorithm, where $N=32$ and $M=64$, respectively. Without loss of generality, the time domain effective channel ${\bf{H}}_{\rm{T}}^{{\rm{eff}}}$ is generated according to~\eqref{Time_domain_channel}, where $P=4$ and the channel coefficients are $[-0.27+0.35i, 0.17+0.01i, 0.56-0.33i,0.31-0.56i]$, the delay indices are $[0, 9, 4, 7]$, and the Doppler indices are $[4.70, -2.26, 1.23, -3.46]$, respectively.
In specific, we consider two constellation mappings at different SNRs, including the quadrature phase shift keying (QPSK) at ${E_s}/{N_0}=12$ dB and 16-quadrature amplitude modulation (16-QAM) at ${E_s}/{N_0}=17$ dB and show state $v_z^{a,{\rm{T}}}\left( {l} \right)$ and the corresponding MSE values in Fig.~\ref{MSE_QPSK_16QAM}.
As observed from the figure, with an increased number of iterations, the
MSE performance of the proposed algorithm first decreases and then saturates at MSEs around $1.3 \times 10^{-4}$ for the QPSK case and around $1.6 \times 10^{-3}$ for the 16-QAM case.
Meanwhile, the derived state evolution shows a close match to the actual MSE performance. This observation indicates that the derived state evolution is consistent with the simulation results.

So far we have investigated the MSE performance of the proposed algorithm via state evolution. In the following subsection, we will focus on the error performance analysis when the proposed algorithm is converged.
\subsection{Analysis of Effective DD Domain SNR}
In this subsection, we will investigate the converged error performance.
In particular, we are interested in the effective DD domain SNR $\eta_{\rm DD}(l)$, because the DD domain detection is a simple component-wise operation, where $\eta_{\rm DD}(l)$ determines the BER performance for a given constellation.
Let us define the second-order time domain OTFS channel matrix ${\bf{G}}_{\rm{T}}^{{\rm{eff}}} \buildrel \Delta \over = {\bf{H}}_{\rm{T}}^{{\rm{eff}}}{\left( {{\bf{H}}_{\rm{T}}^{{\rm{eff}}}} \right)^{\rm{H}}}$.
Noticing that ${{\bf{G}}_{\rm{T}}^{{\rm{eff}}}}$ is a Hermitian matrix by its definition, there exists a unitary matrix $\bf U$ such that ${\bf{G}}_{\rm{T}}^{{\rm{eff}}} = {\bf{U\Lambda }}{{\bf{U}}^{\rm{H}}}$, where ${\bf{\Lambda }}$ is a diagonal matrix whose $(k,k)$-th element is the $k$-th eigenvalue ${\lambda _k}$ of ${\bf{G}}_{\rm{T}}^{{\rm{eff}}}$.
Thus, we can rewrite the \emph{a posteriori} variance $v_z^{p,{\rm{T}}}\left( l \right)$ in~\eqref{a_posteriori_MMSE} according to the eigenvalues ${\lambda _k}$ as shown in the following Lemma.

\textbf{Lemma 1} \emph{(Time domain a posteriori variance)}:
The time domain \emph{a posteriori} variance $v_z^{p,{\rm{T}}}\left( l \right)$ in~\eqref{a_posteriori_MMSE} can be simplified by
\begin{align}
v_z^{p,{\rm{T}}}\left( l \right) = v_z^{{a},{\rm{T}}}\left( l \right) - \frac{{v_z^{{a},{\rm{T}}}\left( l \right)}}{{MN}}\sum\limits_{k = 1}^{MN} {\frac{{v_z^{{a},{\rm{T}}}\left( l \right){\lambda _k}}}{{v_z^{{a},{\rm{T}}}\left( l \right){\lambda _k} + {N_0}}}} . \label{Lemma1}
\end{align}

\emph{Proof}: The proof is given in Appendix C.

In order to obtain some important insights of the effective DD domain SNR $\eta_{\rm DD}(l)$, we need to investigate the property of the eigenvalues ${\lambda _k}$. For the ease of further derivation, let us consider the following assumption.

\textbf{Assumption 2}:
The delay index associated to each resolvable path is different to each other, i.e., ${l_i} \ne {l_j},\forall i \ne j, 1 \le i,j \le P$.

It should be noted that the value of the delay index depends on the specific reflectors corresponding to each resolvable path. Furthermore, in the case where the maximum delay index $l_{\rm max}$ is much larger than $P$, it is unlikely to have a channel realization where different paths share the same delay index. However, later we will show that even without the above assumption, our following derivation can still provide an accurate estimate for the effective DD domain SNR $\eta_{\rm DD}(l)$.
With Assumption 2 in hand, we can derive the following Lemma for ${\bf G}_{\rm T}^{\rm eff}$.

\textbf{Lemma 2} \emph{(Main diagonal element of ${\bf G}_{\rm T}^{\rm eff}$)}:
Under Assumption 2, the main diagonal elements of ${\bf G}_{\rm T}^{\rm eff}$ are of the same value ${\left\| {\bf{h}} \right\|^2}$, where ${\bf{h}} = {\left[ {{h_1},{h_2},...,{h_P}} \right]^{\rm{T}}}$ is the path gain vector.

\emph{Proof}: The Lemma can be straightforwardly derived by noticing that ${{\bm \Pi} ^{{l_i}}}{{\bm\Delta} ^{{k_i} + {\kappa _i}}}{{\bm\Delta} ^{ - {k_i} - {\kappa _i}}}{{\bm\Pi} ^{ - {l_i}}} = {{\bf{I}}_{MN}}$.$\hfill\blacksquare$

According to Lemma 1 and Lemma 2, we can now derive the lower-bound of the DD domain \emph{a priori} variance $v_z^{a,{\rm{DD}}}\left( l \right)$. The corresponding results are summarized in the following Theorem.

\textbf{Theorem 2} \emph{(Lower-bound of $v_z^{a,{\rm{DD}}}\left( l \right)$)}:
Under Assumption 2, the DD domain \emph{a priori} variance $v_z^{a,{\rm{DD}}}\left( l \right)$ is lower-bounded by $\frac{{N_0}}{{{{{\left\| {\bf{h}} \right\|}^2}}}}$, where the lower bound becomes tighter if the time domain \emph{a priori} variance $v_z^{a,{\rm{T}}}\left( l \right)$ tends to zero and the lower bound is achieved when $v_z^{a,{\rm{T}}}\left( l \right)=0$.

\emph{Proof}: The proof is given in Appendix D.

Immediately, we can derive the upper-bound of $\eta_{\rm DD}(l)$ based on Theorem 2.

\textbf{Corollary 1} \emph{(Upper-bound of $\eta_{\rm DD}(l)$)}:
Under Assumption 2, the DD domain effective SNR ${\eta _{\rm{DD}}}\left( {l} \right)$ is upper-bounded by $\frac{{{{\left\| {\bf{h}} \right\|}^2}}}{{{N_0}}}$, where the upper bound becomes tighter with an increased number of iterations.

\emph{Proof}: The proof can be straightforwardly derived from Theorem 2, by noticing that ${\eta _{\rm{DD}}}\left( {l} \right) = \frac{1}{{v_z^{a,{\rm{DD}}}\left( {l} \right)}} \le \frac{{{{\left\| {\bf{h}} \right\|}^2}}}{{{N_0}}}$.$\hfill\blacksquare$

%It is interesting to see from Theorem 2 and Corollary 1 that the proposed algorithm can theoretically approach the error performance of the MRC in the ISI-free case~\cite{tse2005fundamentals}, with a sufficient number of iterations.
%It should be noted that in the ISI-free case, MRC can provide the optimal ML error performance, but the equalization usually requires a prohibitively high complexity that increases exponentially~\cite{tse2005fundamentals}.
It is interesting to see from Theorem 2 and Corollary 1 that the effective DD domain SNR of the proposed algorithm can theoretically approach the maximum receiver SNR for a given fading channel~\cite{tse2005fundamentals}, with a sufficient number of iterations. Equivalently, this observation indicates that the proposed algorithm can approach the error performance of MLSE theoretically given a sufficient number of iterations.
It should be noted that the MLSE can provide the optimal ML error performance, but usually requires a prohibitively high complexity~\cite{tse2005fundamentals}.
Therefore, the proposed algorithm can be viewed as a type of reduced-complexity detection algorithm that can potentially approach the optimal error performance.
On the other hand, we note that the above analysis is based on Assumption 2. For the case where different resolvable paths share the same delay index, we will demonstrate that the
effective DD domain SNR also follows Corollary 1 by numerical simulations.

\begin{figure}
\centering
\includegraphics[width=0.6\textwidth]{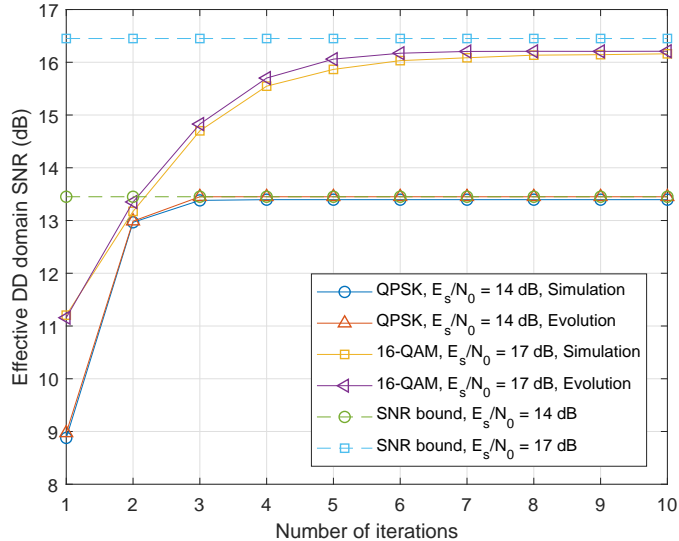}
\caption{Effective DD domain SNRs for OTFS modulation with different number of iterations, where the frame contains $2048$ DD domain QPSK or 16-QAM symbols. In specific, the SNR for the QPSK case is
${E_s}/{N_0}=14$ dB, while that for the 16-QAM case is ${E_s}/{N_0}=17$ dB. The considered wireless channel contains $P=4$ paths with different delay indices for each path. }
\label{SNR_Analysis_different_delay}
\centering
\vspace{-3mm}
\end{figure}

Fig.~\ref{SNR_Analysis_different_delay} shows the effective DD domain SNRs with respect to number of iterations.
Without loss of generality, the time domain effective channel ${\bf{H}}_{\rm{T}}^{{\rm{eff}}}$ is generated according to~\eqref{Time_domain_channel}, where $P=4$ and the channel coefficients are $[-0.04-0.31i, 0.40-0.11i, -0.43+0.18i, 0.59+0.21i]$, the delay indices are $[0, 5, 2, 8]$, and the Doppler indices are $[-3.08, 3.45, -3.94, -0.72]$, respectively.
Similarly, we consider both QPSK and 16-QAM constellations with ${E_s}/{N_0}=14$ dB and ${E_s}/{N_0}=17$ dB, respectively. Meanwhile, we also plot the SNR derived based on the state evolution and the
corresponding SNR upper bound, i.e., $\frac{{{{\left\| {\bf{h}} \right\|}^2}}}{{{N_0}}}$.
As observed from the figure, with an increased number of iterations, the effective DD domain SNR increases.
In specific, the derived SNR based on the state evolution shows a close match to the actual SNR performance based on the simulation.
More importantly, the derived SNR upper bound agrees with the simulation results and state evolution, and the bound becomes tighter as the number of iteration increases, which is consistent with the above analysis.

\begin{figure}
\centering
\includegraphics[width=0.6\textwidth]{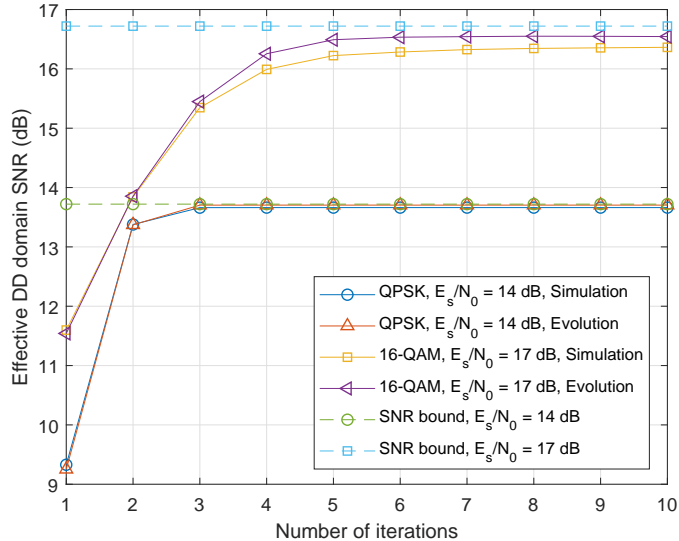}
\caption{Effective DD domain SNRs for OTFS modulation with different number of iterations, where the frame contains $2048$ DD domain QPSK or 16-QAM symbols. In specific, the SNR for the QPSK case is
${E_s}/{N_0}=14$ dB, while that for the 16-QAM case is ${E_s}/{N_0}=17$ dB. The considered wireless channel contains $P=4$ paths and the first two paths share the same delay index.}
\label{SNR_Analysis_same_delay}
\centering
\vspace{-3mm}
\end{figure}

To have a fair comparison, we show the effective DD domain SNRs with respect to the number of iterations with a specific channel realization in Fig.~\ref{SNR_Analysis_same_delay}, where different resolvable paths share the same delay index.
In specific, the channel contains $P=4$ paths and the channel coefficients are $[-0.21+0.27i, 0.17+0.01i, 0.22-0.66i, 0.31-0.46i]$, the delay indices are $[0, 0, 4, 7]$, and the Doppler indices are $[-3.28, 4.45, -1.94, -0.23]$, respectively.
We observe that both QPSK case and 16-QAM case have similar SNR performance to the previous figure. This observation indicates that the above analysis is also valid when different resolvable paths share the same delay index.

We have provided the analysis of the effective DD domain SNR of the proposed algorithm in the convergence. In the following subsection, we will discuss the detection complexity in order to demonstrate the advantage of the proposed algorithm.

\subsection{Analysis of Detection Complexity}
We first consider the computational complexity of module A. It is obvious that the most computation relates to the matrix inverse in~\eqref{MMSE_W}. Generally, the computation complexity order of the matrix inversion is ${\cal O}\left( {{{\left( {MN} \right)}^3}} \right)$.
However, this complexity can be further reduced by considering the specific sparse structure of the time domain effective channel ${\bf{H}}_{\rm{T}}^{{\rm{eff}}}$~\cite{george1988complexity}.
As for module B, it is obvious that the computational complexity is of order ${\cal O}\left( {{MN}} \right)$ because of the component-wise operation. On the other hand,
the computational complexity for the domain transformation can be low by considering the special structure of the corresponding kernels. In particular, the Kronecker products ${\bf{F}}_N \otimes {{\bf{I}}_M}$ and ${\bf{F}}_N^{\rm{H}} \otimes {{\bf{I}}_M}$ in~\eqref{m_x_pri} and~\eqref{m_z_post_B} can be efficiently calculated based on FFT and IFFT and the corresponding computation complexity will be ${\cal O}\left( {MN\log N} \right)$.
Therefore, we can calculate the total detection complexity of the proposed algorithm per iteration, which is given by ${\cal O}\left( {{{\left( {MN} \right)}^3} + 2MN\log N + MN} \right)$.
%Note that the above detection complexity can be viewed as an upper bound for the actual detection complexity, which does not take into account of the sparsity of ${\bf{H}}_{\rm{T}}^{{\rm{eff}}}$.
%Furthermore,
It should be noted that the detection complexity does not increase in the presence of the fractional Doppler indices. In comparison of the proposed algorithm, the detection complexity of the optimal MLSE detection is exponential to the number of non-zero elements per row/column of the corresponding channel matrix, which can be as high as ${\cal O}\left( {{\cal X}{{\left( {\mathbb A} \right)}^{MN}}} \right)$ when the fractional Doppler shift exists.
Therefore, the proposed algorithm
can significantly reduce the detection complexity compared to the optimal MLSE detection, especially in the presence of fractional Doppler shifts.

%For reference, in the case of QPSK mapping, where $M=4$ and $N=4$, the required computational complexity of the proposed algorithm can be
%can be ${\cal O}\left( 4096\right)$ for MMSE detection, $4.2 \times 10^9$ for MLSE detection, and $4176$ for the proposed algorithm, respectively. It is not hard to observe that the proposed algorithm can provide significant complexity reduction compared to the MLSE detection.

%\begin{table*}[htbp]
%\caption{Computational Complexity Comparison}
%\centering
%\begin{tabular}{|c|c|c|c|c|c|c|c|}
%\hline
%Detection method~&~Complexity (per iteration)\\
%\hline
%MMSE~&~${\cal O}\left( {{{\left( {MN} \right)}^3}} \right)$\\
%\hline
%MLSE~&~${\cal O}\left( {{\cal X}{{\left( {\mathbb A} \right)}^{MN}}} \right)$\\
%\hline
%Proposed~&~${\cal O}\left( {{{\left( {MN} \right)}^3} + 2MN\log N + MN} \right)$\\
%\hline
%\end{tabular}
%\vspace{-3mm}
%\label{Complexity_Compare}
%\end{table*}
%
%
%We have provided a comprehensive analysis of the proposed algorithm in terms of the error performance and detection complexity. In the next section, we will demonstrate the BER performance of the proposed algorithm by numerical simulations.

\section{Numerical Results}
In this section, we will evaluate the BER performance of the proposed algorithm. We consider the average BER performance with a sufficient number of realizations of the channel. Specifically, the corresponding channel matrix is generated according to~\eqref{Time_domain_channel}, the channel coefficients are randomly generated based on a uniform power delay profile, and the delay and Doppler indices are randomly generated within the range of $[0, l_{\rm max}]$ and $[-k_{\rm max}, k_{\rm max}]$, where $l_{\rm max}=10$ and $k_{\rm max}=5$. We note that, as mentioned before, the delay index can only be an integer number, while the Doppler index can be a fractional number~\cite{Raviteja2018interference}.
Without loss of generality, we consider the QPSK modulated OTFS system with different number of paths.
We also provide other detection methods for comparison that include the MMSE detection based on the DD domain effective channel ${\bf H}_{\rm DD}^{\rm eff}$,
DD domain detection based on the sum-product algorithm (SPA)~\cite{li2020hybrid}, and the DD domain message passing algorithm in~\cite{Raviteja2018interference}.
The considered SPA detection is derived based on the graphical model corresponding to the DD domain effective channel, whose computational complexity 
can be ${\cal O}\left( {{\cal X}{{\left( {\mathbb A} \right)}^{MN}}} \right)$ in the case of complex fractional Doppler shifts~\cite{li2020hybrid}. In particular, the 
considered SPA detection can theoretically approach the error performance of the optimal MLSE detection and achieve the same performance when the graphical model does not contain any cycle~\cite{li2020hybrid}.
However, since the DD domain SPA detection requires a very high detection complexity in the fractional Doppler case, we only consider the integer Doppler case for simplicity.

\begin{figure}
\centering
\includegraphics[width=0.6\textwidth]{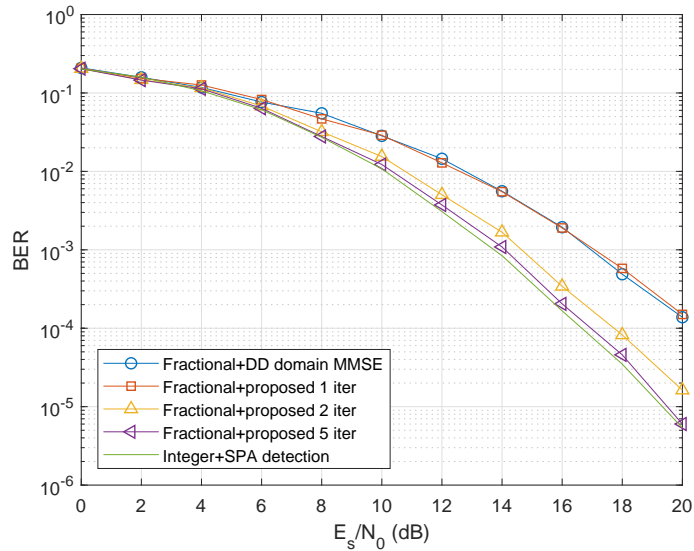}
\caption{BER performance for OTFS modulation with $P=4$, where the error performance of the proposed algorithm is compared with that of DD domain MMSE detection and MLSE detection with integer Doppler shifts.}
\label{P4_BER}
\centering
\end{figure}
Fig.~\ref{P4_BER} shows the BER performance with $P=4$.
As observed from the figure, the error performance of the proposed algorithm with one iteration is almost the same as that with the MMSE detection. This observation indicates that applying the same detection method for OTFS modulation over different domains can result in the similar error performance.
Furthermore,
with the increased number of iterations, the proposed algorithm outperforms the MMSE detection and almost achieves the performance of SPA detection with only integer Doppler indices.
In specific, with BER below $1 \times 10^{-4}$, the proposed algorithm with $5$ iterations has only a marginal performance gap (around $0.2$ dB) compared to the SPA detection.
This observation clearly substantiates our theoretical analysis in the previous section.

\begin{figure}
\centering
\vspace{-5mm}
\includegraphics[width=0.6\textwidth]{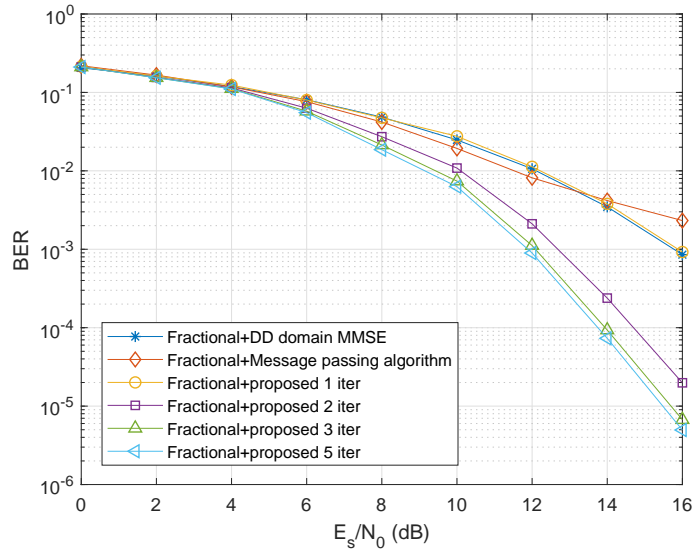}
\caption{BER performance for OTFS modulation with fractional Doppler shifts, where $P=10$.}
\label{P10_BER}
\centering
\end{figure}
Fig.~\ref{P10_BER} shows the BER performance with $P=10$. Under such a complex channel condition, the DD domain effective channel ${\bf H}_{\rm DD}^{\rm eff}$ can be very dense and conventional detection methods based on ${\bf H}_{\rm DD}^{\rm eff}$ may not have a good error performance.
As shown in the figure, the proposed algorithm with one iteration performs almost the same with the DD domain MMSE detection, which is consistent with the observation in the previous figure. Meanwhile, the DD domain message passing algorithm~\cite{Raviteja2018interference} also shows a similar error performance to both the DD domain MMSE detection and the proposed algorithm with one iteration.
However, with the increased number of iterations, the proposed algorithm significantly outperforms the MMSE detection and message passing algorithm. Specifically, at BER $ \approx 1 \times {10^{ - 3}}$, with only $2$ iterations, the error performance of the proposed algorithm shows an around $3.2$ dB gain compared to that of the MMSE detection and the gain to the message passing algorithm is even more.
Furthermore, with $5$ iterations, the SNR gain to the MMSE detection performance is increased to around $4.1$ dB. This observation shows the advantage of the proposed algorithm over conventional detection algorithms, which also agrees with our previous theoretical analysis.

\section{Conclusion}
In this paper, we proposed a novel cross domain iterative detection for OTFS modulation. We derived the state evolution for the proposed algorithm and investigated its detection performance including both the MMSE performance and the DD domain effective SNR. In particular, we show that the proposed algorithm can approach the error performance of 
MLSE detection even in the presence of complex fractional Doppler shifts, but only requires a much lower detection complexity. Our analytical results are explicitly verified by simulation results, where a significant performance improvement can be observed compared to the conventional detection algorithms for channels with fractional Doppler shifts.
The cross domain signal processing may be a new research direction for OTFS modulation and general multicarrier modulation schemes.
Our future work will investigate the cross domain channel estimation schemes and cross domain precoding schemes. 
\appendices
\section{Justification of Problem Formulation~\eqref{DD_detection_from_TD}}
Note that the DD domain detection is carried out based on the estimates of time domain signal.
As a common approach, we model the extrinsic mean of $\bf z$ by
\begin{align}
{\bf{m}}_{\bf{z}}^{e,{\rm{T}}} ={\bf{z}}+ {{\bf{\hat w}}}, \label{Time_domain_estimation}
\end{align}
where ${{\bf{\hat w}}}$ is a white Gaussian noise sample vector with zero mean and a diagonal covariance matrix ${\bf{C}}_{\bf{z}}^{a,{\rm{DD}}}={\bf{C}}_{\bf{z}}^{e,{\rm{T}}}$.
In particular, the term ${\bf{\hat w}}$ stands for the inaccuracy of the MMSE estimation in the time domain. %However, such an inaccuracy can be difficult to deal with by the time domain MMSE estimation solely. Therefore, we can convert the time domain estimates to the DD domain and exploit the discrete constellation constraint of the DD domain symbols, in order to improve the error performance.
%However, this issue must be solved with care. In what follows, we will discuss possible solutions for this issue and justify our DD domain detection problem formulation~\eqref{DD_detection_from_TD}.
In order to carry out the DD domain symbol detection, a straightforward way is to first convert the extrinsic information ${\bf{m}}_{\bf{z}}^{e,{\rm{T}}}$ of the time domain signal $\bf z$ to DD domain and then compare it with the DD domain symbol constellation.
In specific, we can directly apply the transformation ${{{\bf{F}}_N} \otimes {{\bf{I}}_M}}$ onto~\eqref{Time_domain_estimation} and obtain the corresponding DD domain symbol estimates, i.e.,
\begin{align}
{\bf{m}}_{\bf{x}}^{a,{\rm{DD}}}= \left( {{{\bf{F}}_N} \otimes {{\bf{I}}_M}} \right){\bf{m}}_{\bf{z}}^{e,{\rm{T}}}={\bf{x}} + \left( {{{\bf{F}}_N} \otimes {{\bf{I}}_M}} \right){{\bf{\hat w}}},
\end{align}
where the term $\left( {{{\bf{F}}_N} \otimes {{\bf{I}}_M}} \right){{\bf{\hat w}}}$ is the equivalent noise in the DD domain and its covariance matrix is denoted by ${\bf{C}}_{\bf{x}}^{a,{\rm{DD}}}$.
Recalling~\eqref{T_transmitted_symbol_vec}, we have
\begin{align}
{\bf{C}}_{\bf{x}}^{a,{\rm{DD}}} &= \left( {{{\bf{F}}_N} \otimes {{\bf{I}}_M}} \right){\bf{C}}_{\bf{z}}^{a,{\rm{DD}}}\left( {{\bf{F}}_N^{\rm{H}} \otimes {{\bf{I}}_M}} \right) .\label{c_x_pri}
\end{align}
It should be noted that in the asymptotical regime, i.e., $MN$ tends to infinity, the diagonal entries of the diagonal matrix ${\bf{C}}_{\bf{z}}^{a,{\rm{DD}}}$ tend to be of the same value owing to the law of large numbers, i.e., ${\bf{C}}_{\bf{z}}^{a,{\rm{DD}}} \propto {{\bf{I}}_{MN}}$. In this case, we have ${\bf{C}}_{\bf{x}}^{a,{\rm{DD}}}={\bf{C}}_{\bf{z}}^{a,{\rm{DD}}}$ and the DD domain symbol detection can be done in a straightforward symbol-by-symbol fashion. However, in practice, the diagonal entries of ${\bf{C}}_{\bf{z}}^{a,{\rm{DD}}}$ may not be of the same value due to the specific noise values.
In this case, ${\bf{C}}_{\bf{x}}^{a,{\rm{DD}}}$ can be non-diagonal and dense, which not only conflicts with the i.i.d. assumption of the entries in $\bf x$, but also potentially undermines the detection performance.
In contrast to the above solution, we can consider the DD domain detection problem formulation in~\eqref{DD_detection_from_TD}. As can be noticed from Section III-C, the proposed DD domain detection algorithm based on~\eqref{DD_detection_from_TD} bypasses the problem that the diagonal entries of ${\bf{C}}_{\bf{z}}^{a,{\rm{DD}}}$ are of different values and can still be efficiently carried out in a symbol-by-symbol fashion.
%Based on the above discussion, the problem formulation~\eqref{DD_detection_from_TD} is justified.
\section{Proof of Proposition 1}
According to the definition of extrinsic information, we can derive the extrinsic mean $m_{\bf{x}}^{e,{\rm{DD}}}\left[ k \right]$ by excluding the contribution of $x[k]$ in~\eqref{m_x_post_B_element}.
By noticing that $m_{\bf{x}}^{p,{\rm{DD}}}\left[ k \right] = {\mathbb E}\left[ {x\left[ k \right]|{\bf{m}}_{\bf{z}}^{e,{\rm{T}}}} \right] = {\mathbb E}\left[ {x\left[ k \right]|{\bf{m}}_{\bf{x}}^{a,{\rm{DD}}}} \right]$, we have
\begin{align}
m_{\bf{x}}^{e,{\rm{DD}}}\left[ k \right] = {\mathbb E}\left[ {x\left[ k \right]|{\bf{m}}_{\bf{x}}^{a,{\rm{DD}}}/m_{\bf{x}}^{a,{\rm{DD}}}\left[ k \right]} \right],
\end{align}
where ${{\bf{m}}_{\bf{x}}^{a,{\rm{DD}}}/m_{\bf{x}}^{a,{\rm{DD}}}\left[ k \right]}$ denotes ${\bf m}_{\bf{x}}^{e,{\rm{DD}}}$ excluding the entry of ${m_{\bf{x}}^{a,{\rm{DD}}}\left[ k \right]}$.
Furthermore, it can be noticed that the \emph{a posteriori} probability of $x[k]$ only relates to ${m_{\bf{x}}^{a,{\rm{DD}}}\left[ k \right]}$ instead of other entries in ${\bf m}_{\bf{x}}^{e,{\rm{DD}}}$.
Therefore, it can be shown that $m_{\bf{x}}^{e,{\rm{DD}}}\left[ k \right] =0$~\cite{Ma2015Turbo}. Similarly, it can be shown that DD domain detection cannot provide any extrinsic information in terms of the covariance matrix either, due to the component-wise operation~\cite{Ma2015Turbo}.
This completes the proof of Proposition 1.$\hfill\blacksquare$
\section{Proof of Lemma 2}
According to~\eqref{a_posteriori_MMSE}, we have
\begin{align}
&v_z^{{p},{\rm{T}}}\left( l \right)\notag\\
=&v_z^{{a},{\rm{T}}}\left( l \right) - \frac{{{{\left( {v_z^{{a},{\rm{T}}}\left( l \right)} \right)}^2}}}{{MN}}{\rm{Tr}}\left( {{{\left( {{\bf{H}}_{\rm{T}}^{{\rm{eff}}}} \right)}^{\rm{H}}}{{\left( {v_z^{{a},{\rm{T}}}\left( l \right){\bf{H}}_{\rm{T}}^{{\rm{eff}}}{{\left( {{\bf{H}}_{\rm{T}}^{{\rm{eff}}}} \right)}^{\rm{H}}} + {N_0}{{\bf{I}}_{MN}}} \right)}^{ - 1}}{\bf{H}}_{\rm{T}}^{{\rm{eff}}}} \right) \notag\\
=&v_z^{{a},{\rm{T}}}\left( l \right) - \frac{{{{\left( {v_z^{{a},{\rm{T}}}\left( l \right)} \right)}^2}}}{{MN}}{\rm{Tr}}\left( {{{\left( {v_z^{{a},{\rm{T}}}\left( l \right){\bf{G}}_{\rm{T}}^{{\rm{eff}}} + {N_0}{{\bf{I}}_{MN}}} \right)}^{ - 1}}{\bf{G}}_{\rm{T}}^{{\rm{eff}}}} \right). \label{Lemma2_der1}
\end{align}
Furthermore, by considering ${\bf{G}}_{\rm{T}}^{{\rm{eff}}} = {\bf{U\Lambda }}{{\bf{U}}^{\rm{H}}}$, ~\eqref{Lemma2_der1} can be further simplified by
\begin{align}
v_z^{{p},{\rm{T}}}\left( l \right)
=&v_z^{{a},{\rm{T}}}\left( l \right) - \frac{{{{\left( {v_z^{{a},{\rm{T}}}\left( l \right)} \right)}^2}}}{{MN}}{\rm{Tr}}\left( {{{\left( {v_z^{{a},{\rm{T}}}\left( l \right){\bf{U\Lambda }}{{\bf{U}}^{\rm{H}}} + {N_0}{{\bf{I}}_{MN}}} \right)}^{ - 1}}{\bf{U\Lambda }}{{\bf{U}}^{\rm{H}}}} \right)\notag\\
=&v_z^{{a},{\rm{T}}}\left( l \right) - \frac{{{{\left( {v_z^{{a},{\rm{T}}}\left( l \right)} \right)}^2}}}{{MN}}{\rm{Tr}}\left( {{\bf{U}}{{\left( {v_z^{{a},{\rm{T}}}\left( l \right){\bf{\Lambda }} + {N_0}{{\bf{I}}_{MN}}} \right)}^{ - 1}}{\bf{\Lambda }}{{\bf{U}}^{\rm{H}}}} \right)\notag\\
=&v_z^{{a},{\rm{T}}}\left( l \right) - \frac{{v_z^{{a},{\rm{T}}}\left( l \right)}}{{MN}}\sum\limits_{k = 1}^{MN} {\frac{{v_z^{{a},{\rm{T}}}\left( l \right){\lambda _k}}}{{v_z^{{a},{\rm{T}}}\left( l \right){\lambda _k} + {N_0}}}} .
\end{align}
This completes the proof of Lemma 2.$\hfill\blacksquare$
\section{Proof of Theorem 2}
The proof is based on the application of Jensen's inequality. Let us consider the following function $f\left( \lambda  \right) = {\frac{{v\lambda }}{{v\lambda  + {N_0}}}}$,
whose second-order derivative with respect to $\lambda$ is of the form ${f^{''}}\left( \lambda  \right) =  - \frac{{2{v^2}{N_0}}}{{{{\left( {v\lambda  + {N_0}} \right)}^3}}}$.
It can be shown that with $v$ and $N_0$ strictly above zero, the above function is a concave function.
Furthermore, with $v$ close to zero, ${f^{''}}\left( \lambda  \right)$ tends to be a zero, which indicates that $f\left( \lambda  \right)$ tends to be a linear function.
Therefore, according to Jensen's inequality, it is obvious that
\begin{align}
\frac{1}{{MN}}\sum\limits_{k = 1}^{MN} {f\left( {{\lambda _k}} \right)}  \le f\left( {\frac{1}{{MN}}\sum\limits_{k = 1}^{MN} {{\lambda _k}} } \right) = \frac{{\frac{v}{{MN}}\sum\limits_{k = 1}^{MN} {{\lambda _k}} }}{{\frac{v}{{MN}}\sum\limits_{k = 1}^{MN} {{\lambda _k}}  + {N_0}}},
\end{align}
where the bound becomes tighter with decreasing $v$ and the equality is achieved when $v=0$.
Therefore, we have
\begin{align}
v_z^{p,{\rm{T}}}\left( l \right) &= v_z^{{a},{\rm{T}}}\left( l \right) -  \frac{{v_z^{{a},{\rm{T}}}\left( l \right)}}{{MN}}\sum\limits_{k = 1}^{MN} {\frac{{v_z^{{a},{\rm{T}}}\left( l \right){\lambda _k}}}{{v_z^{{a},{\rm{T}}}\left( l \right){\lambda _k} + {N_0}}}} \notag\\
&\ge v_z^{{a},{\rm{T}}}\left( l \right) - \frac{{{{\left( {v_z^{{a},{\rm{T}}}\left( l \right)} \right)}^2} \frac{1}{{MN}}\sum\limits_{k = 1}^{MN} {{\lambda _k}} }}{{v_z^{{a},{\rm{T}}}\left( l \right) \frac{1}{{MN}}\sum\limits_{k = 1}^{MN} {{\lambda _k}}  + {N_0}}}, \label{Theorem2_der1}
\end{align}
where the lower bound becomes tighter with the decrease of $v_z^{{a},{\rm{T}}}\left( l \right)$ and the equality is achieved when $v_z^{{a},{\rm{T}}}\left( l \right)$ becomes zero.
Notice that $\sum\limits_{k = 1}^{MN} {{\lambda _k}}  = {\rm{Tr}}\left( {{\bf{G}}_{\rm{T}}^{{\rm{eff}}}} \right)$. Thus, by considering Lemma 2,~\eqref{Theorem2_der1} becomes
\begin{align}
v_z^{{p},{\rm{T}}}\left( l \right) \ge v_z^{{a},{\rm{T}}}\left( l \right) - \frac{{{{\left( {v_z^{{a},{\rm{T}}}\left( l \right)} \right)}^2}{{\left\| {\bf{h}} \right\|}^2}}}{{v_z^{{a},{\rm{T}}}\left( l \right){{\left\| {\bf{h}} \right\|}^2} + {N_0}}}. \label{Theorem2_der2}
\end{align}
By substituting~\eqref{Theorem2_der2} into~\eqref{state_evolution_A_B}, after some manipulations, we arrive at
\begin{align}
v_z^{{a},{\rm{DD}}}\left( l \right) &= \frac{1}{{\frac{1}{{v_z^{{p},{\rm{T}}}\left( l \right)}} - \frac{1}{{v_z^{{a},{\rm{T}}}\left( l \right)}}}} \ge \frac{1}{{\frac{1}{{v_z^{{a},{\rm{T}}}\left( l \right) - \frac{{v_z^{{a},{\rm{T}}}\left( l \right){{\left\| {\bf{h}} \right\|}^2}}}{{v_z^{{a},{\rm{T}}}\left( l \right){{\left\| {\bf{h}} \right\|}^2} + {N_0}}}}} - \frac{1}{{v_z^{{a},{\rm{T}}}\left( l \right)}}}}\notag\\
&= \frac{{v_z^{{a},{\rm{T}}}\left( l \right) - \frac{{{{\left( {v_z^{{a},{\rm{T}}}\left( l \right)} \right)}^2}{{\left\| {\bf{h}} \right\|}^2}}}{{v_z^{{a},{\rm{T}}}\left( l \right){{\left\| {\bf{h}} \right\|}^2} + {N_0}}}}}{{\frac{{v_z^{{a},{\rm{T}}}\left( l \right){{\left\| {\bf{h}} \right\|}^2}}}{{v_z^{{a},{\rm{T}}}\left( l \right){{\left\| {\bf{h}} \right\|}^2} + {N_0}}}}} = \frac{{{N_0}}}{{{{\left\| {\bf{h}} \right\|}^2}}}.
\end{align}
This completes the proof of Theorem 2.$\hfill\blacksquare$
%In specific, the term $v_z^{{\rm{post,A}}}$ in~\eqref{Fixed_point_der1} denotes the average \emph{a posteriori} variance of the time domain MMSE estimation when the algorithm is converged, which is given by $v_z^{{\rm{post,A}}} = f\left( {\frac{1}{{MMSE\left( {{\eta _ * }} \right)}} - {\eta _ * }} \right)$, where
%\begin{align}
%f\left( v \right) = v - \mathop {\lim }\limits_{MN \to \infty } \frac{{{v^2}}}{{MN}}{\rm{Tr}}\left( {{{\left( {{\bf{H}}_{\rm{T}}^{{\rm{eff}}}} \right)}^{\rm{H}}}{{\left( {v{\bf{H}}_{\rm{T}}^{{\rm{eff}}}{{\left( {{\bf{H}}_{\rm{T}}^{{\rm{eff}}}} \right)}^{\rm{H}}} + {N_0}{{\bf I}_{MN}}} \right)}^{ - 1}}{\bf{H}}_{\rm{T}}^{{\rm{eff}}}} \right).
%\end{align}
%After some straightforward manipulations,~\eqref{Fixed_point_der1} becomes
%\begin{align}
%MMSE\left( {{\eta _*}} \right){\rm{ = }}v_z^{{\rm{post,A}}}.
%\end{align}
%In specific, this equation indicates that both the

\bibliographystyle{IEEEtran}
\bibliography{OTFS_references}
\end{document}